\documentclass{aastex631}

\shorttitle{SGR 1935+2154 at 110 MHz } \shortauthors{Hu et al.}
\graphicspath{{./}{figures/}}
\usepackage{CJK}
\usepackage[T2A]{fontenc}
\usepackage[utf8]{inputenc}
\usepackage{amsmath}

\usepackage{longtable}
\usepackage{booktabs}
\usepackage{hyperref}
\usepackage[all]{hypcap}
\usepackage{CJK}
\makeatletter

\newcommand{\Rmnum}[1]{\expandafter\@slowromancap\romannumeral #1@}
\makeatother

\begin{document}
\begin{CJK*}{UTF8}{gbsn}

\title{Constraints on the Low-frequency Radio Emission of the Galactic FRB Source SGR 1935+2154}

\author[0000-0002-5238-8997]{Chen-Ran Hu (胡宸然)}
\affiliation{School of Astronomy and Space Science, Nanjing
University, Nanjing 210023, China}

\author[0009-0000-7501-2215]{Jinhuang Cao (曹锦煌)}
\affiliation{National Astronomical Observatories, Chinese Academy of
Sciences, 20A Datun Road, Chaoyang District, Beijing 100101, China}
 \affiliation{University of Chinese Academy of Sciences, Beijing 100049,
China}

\author[0000-0003-0042-0884]{S.A. Tyul'bashev}
\thanks{Email: serg@prao.ru}
\affiliation{Lebedev Physical Institute of Astro Space Center of
Pushchino Radio Astronomy Observatory, PRAO, Radiotelescopnaya 1a, Moscow
region, 142290, Russia}

\author[0000-0002-3386-7159]{Pei Wang (王培)}
\thanks{Email: wangpei@nao.cas.cn}
\affiliation{State Key Laboratory of Radio Astronomy and Technology, 
NAOC, Chinese Academy of Sciences, Beijing 100101, China}
 \affiliation{Institute for Frontiers in Astronomy and Astrophysics, Beijing
Normal University, Beijing 102206, China}

\author[0000-0001-7199-2906]{Yong-Feng Huang (黄永锋)}
\thanks{Email: hyf@nju.edu.cn}
\affiliation{School of Astronomy and Space Science, Nanjing
University, Nanjing 210023, China}
 \affiliation{Key Laboratory of
Modern Astronomy and Astrophysics (Nanjing University), Ministry
of Education, China}

\author[0009-0005-9243-7633]{E.A. Brylyakova}
\affiliation{Lebedev Physical Institute of Astro Space Center of
Pushchino Radio Astronomy Observatory, PRAO, Radiotelescopnaya 1a, Moscow
region, 142290, Russia}

\author[0000-0001-6684-989X]{G.E. Tyul'basheva}
\affiliation{Institute of Mathematical Problems of Biology, Branch of
Keldysh Institute of Applied Mathematics of Russian Academy of Sciences,
IMPB RAS, Professor Vitkevich 1, Moscow region, 142290, Russia}

\author[0000-0001-9648-7295]{Jin-Jun Geng (耿金军)}
\affiliation{Purple Mountain Observatory, Chinese Academy of
Sciences,  Nanjing 210023, China}

\author[0000-0003-3230-7587]{Orkash Amat (吾热卡西·艾麦提)}
\affiliation{School of Astronomy and Space Science, Nanjing
University,  Nanjing 210023,  China}

\author[0000-0002-6189-8307]{Ze-Cheng Zou (邹泽城)}
\affiliation{School of Astronomy and Space Science, Nanjing
University, Nanjing 210023, China}

\author[0009-0002-8460-1649]{Chen Du (杜琛)}
\affiliation{School of Astronomy and Space Science, Nanjing
University, Nanjing 210023,  China}

\author[0000-0001-9227-3716]{Nurimangul Nurmamat (努尔曼古丽·努尔麦麦提)}
\affiliation{Guangxi Key Laboratory for Relativistic Astrophysics, 
School of Physical Science and Technology, Guangxi University, Nanning 
530004, China}

\author[0000-0003-0721-5509]{Lang Cui (崔朗)}
\affiliation{State Key Laboratory of Radio Astronomy and Technology, 
Xinjiang Astronomical Observatory, CAS, 150 Science 1-Street, Urumqi, 
Xinjiang, 830011, China}
 \affiliation{Xinjiang Key Laboratory of Radio Astrophysics, 150 Science
1-Street, Urumqi 830011, China}

\author[0000-0001-7943-4685]{Fan Xu (许帆)}
\affiliation{Institute of Space Weather, School of Atmospheric Physics,
Nanjing University of Information Science and Technology, Nanjing 210044,
China}

\author[0009-0000-0467-0050]{Xiao-Fei Dong (董小飞)}
\affiliation{School of Astronomy and Space Science, Nanjing
University, Nanjing 210023,  China}

\author[0000-0002-2191-7286]{Chen Deng (邓晨)}
\affiliation{School of Astronomy and Space Science, Nanjing
University, Nanjing 210023, China}

\begin{abstract}

We present a search for radio pulses from the Galactic magnetar
SGR 1935+2154, a well-known source of fast radio bursts (FRBs), at
$\sim$110 MHz using the Large Phased Array (LPA) of the Pushchino
Radio Astronomy Observatory. Data from two active periods in 2020
(March -- May and September -- November, with $\sim 3.5$ minutes
of daily coverage) were analyzed with new methods tailored to both
FRB-like single pulses and pulsar-like periodic signals. No
significant FRB-like pulses were found. Using Monte Carlo
simulations, $3\sigma$ upper limits were derived for the burst
rate: for a log-normal energy distribution the limit is
$\sim$${10}^{1.5}~{\rm{d}}^{-1}$ for a mean of average
monochromatic isotropic luminosity $L_{\nu{\rm
,mean}}\sim1.3\times{10}^{29}~{\rm{erg~s^{-1}~ {Hz}^{-1}}}$ and a
natural log-space scatter of
$\sigma\sim0.85$; while for a power-law distribution it is
$\sim$${10}^{1.8}~{\rm{d}}^{-1}$ for an index $\beta\lesssim3.0$
and a minimum average monochromatic isotropic luminosity
$L_{\nu{\rm{,min}}}\lesssim0.7\times{10}^{25}~{\rm{erg~s^{-1}~{Hz}^{-1}}}$.
When folded at the known 3.24781628 s period of SGR 1935+2154, a
weak pulse was noted (S/N $<$ 3.16), but the significance is
insufficient for a secure detection of the pulsar-like emission
signal. A conservative upper limit on the average monochromatic
isotropic luminosity of any possible periodic emission is
$2.08\times{10}^{19}~{\rm{erg~s^{-1}~{Hz}^{-1}}}$. Our results
offer meaningful low-frequency upper limits on the burst rate of
SGR 1935+2154, and hint for very faint pulsar-like radiation at
meter wavelengths.


\end{abstract}



\section{Introduction}
\label{sec1:introduction}

Fast radio bursts (FRBs) are millisecond-duration transients of
cosmological origin that release enormous energy in the radio band
\citep{2019ARA&A..57..417C, 2019A&ARv..27....4P, 2022A&ARv..30....2P}.
Their progenitors and emission mechanisms remain highly debated
\citep{2019PhR...821....1P, 2020Natur.587...45Z, 2021SCPMA..6449501X,
2023RvMP...95c5005Z, 2023ApJS..269...17H}, partly because multi-wavelength
counterparts are rarely detected.

A number of repeating FRB sources have been observed across
various radio sub-bands (from $\sim$0.1 GHz to $\sim$8 GHz).
Well-known examples include FRB 20121102A
\citep{2016Natur.531..202S, 2018Natur.553..182M,
2018ApJ...863....2G, 2018ApJ...863..150S, 2018ApJ...866..149Z,
2020MNRAS.495.3551R, 2021ApJ...908L..10H, 2021Natur.598..267L,
2022MNRAS.515.3577H, 2023MNRAS.519..666J}, FRB 20180301A
\citep{2019MNRAS.486.3636P, 2020Natur.586..693L,
2022ApJ...930..172L, 2023MNRAS.526.3652K}, FRB 20180916B
\citep{2020Natur.577..190M, 2020Natur.582..351C,
2020ApJ...896L..40P, 2020MNRAS.499L..16M, 2021Natur.596..505P,
2021ApJ...911L...3P, 2021ApJS..257...59C, 2022ApJ...932...98S,
2023ApJ...950...12M, 2023MNRAS.524.3303B, 2026ApJS..283...34C},
FRB 20190520B \citep{2022Natur.606..873N, 2022Sci...375.1266F,
2023Sci...380..599A, 2026SciBu..71...76N}, FRB 20201124A
\citep{2021ApJS..257...59C, 2021MNRAS.508.5354H,
2022MNRAS.509.2209M, 2022ApJ...927L...3N, 2022ApJ...927...59L,
2022MNRAS.512.3400K, 2022Natur.609..685X, 2022RAA....22l4002Z,
2025ApJ...982..154N, 2026ApJS..283...34C}, and FRB 20240114A
\citep{2024MNRAS.533.3174T, 2024ApJ...977..177K,
2025arXiv250714707Z, 2025RAA....25h5009H, 2025ApJS..278...49X,
2025ApJ...989...15P, 2026ApJ...997..334S, 2026ApJ...998..276Z,
2026SCPMA..6949512Z, 2026arXiv260320663W}, all of which are
extragalactic. Several groups have summarized the burst frequency
distributions of these repeaters \citep{2023MNRAS.522.5600L,
2024ApJ...966..115L}. However, the multi-band radio observations
are not simultaneous. Attempts to obtain simultaneous detections
at two or more radio sub-bands have largely failed
\citep{2021ApJ...911L...3P, 2021ATel14605....1K,
2021Natur.596..505P, 2025ApJ...992..185W}. For a long time, the only
extragalactic source that had been simultaneously detected by multiple
radio telescopes was FRB 20121102A, the first identified repeater, with
the longest monitoring history. It was found to produce bursts at
1.10--1.80 GHz (the Nan\c{c}ay radio telescope) and 0.90--1.67 GHz
(the MeerKAT radio telescope) \citep{2020MNRAS.496.4565C}, as well
as at 1.15--1.73 GHz (the 305-m William E. Gordon Telescope at the
Arecibo Observatory, Arecibo) and 2.50--3.50 GHz (the Karl G.
Jansky Very Large Array, VLA) \citep{2017ApJ...850...76L}.
Nevertheless, the simultaneous burst frequency ranges are
relatively close; recently, simultaneous bursts have also been reported
for the currently most active repeater FRB 20240114A, though all such
bursts fell within the L band \citep{2026arXiv260518513O}, suggesting that
repeating FRBs tend to be active only within a narrow frequency band at
any given epoch. This may be linked to their intrinsically narrow,
Gaussian-like spectral shape, whose physical origin remains unclear
\citep{2025arXiv250318084H}.

Searches for multi-wavelength counterparts in X-rays, gamma-rays
\citep{2020ApJ...901..165S, 2021A&A...656L..15P,
2022ApJ...930..172L, 2025A&A...695L..10E, 2017ApJ...846...80S},
and even in the optical band \citep{2023A&A...676A..17T,
2025A&A...704A..25G} have all yielded nondetections. Nonetheless,
these upper limits provide valuable constraints on the
multi-wavelength energetics of FRB counterparts.

The scarcity of multi-wavelength counterparts may simply reflect
that most FRBs are too distant and their counterparts are too
faint to be detectable with current instruments. Fortunately, we
have a Galactic FRB source, the magnetar SGR 1935+2154 (hereafter
SGR 1935). It produced the well-known FRB 200428
\citep{2020Natur.587...54C, 2020Natur.587...59B} accompanied by a
series of X-ray bursts \citep{2020ApJ...898L..29M,
2021NatAs...5..378L, 2021NatAs...5..372R, 2021NatAs...5..401T}.
Importantly, FRB 200428 was simultaneously observed at different
radio frequencies: 0.40--0.80 GHz by the Canadian Hydrogen
Intensity Mapping Experiment telescope (CHIME)
\citep{2020Natur.587...54C} and 1.28--1.47 GHz by the Survey for
Transient Astronomical Radio Emission 2 array (STARE2)
\citep{2020Natur.587...59B}. In the second month after FRB 200428,
i.e., May 2020 (UTC time, same hereafter), the Five-hundred-meter
Aperture Spherical radio Telescope (FAST) detected a burst from
SGR 1935 at 1.00--1.50 GHz \citep{2020ATel13699....1Z}, followed
by two bursts detected with a single 25-m Westerbork dish (RT-1)
at 1.26--1.39 GHz \citep{2021NatAs...5..414K}. Half a year later,
in October 2020, three radio bursts from SGR 1935 were detected
again by CHIME \citep{2020ATel14074....1G}, followed by nearly a
month of intense pulsar-like (periodic) emission observed by FAST
\citep{2023SciA....9F6198Z, 2024ApJS..275...39W}. X-ray
observations revealed that this active episode was accompanied by
sudden spin changes, i.e., one anti-glitch (spin-down) on 5
October \citep{2023NatAs...7..339Y} and two glitches (spin-up) on
14 October \citep{2024Natur.626..500H}.

These multi-wavelength detections of SGR 1935 have greatly
enriched our understanding of FRB physics, including the emission
mechanism, the production of high-energy counterparts
\citep{2020MNRAS.498.1397L, 2020ApJ...900L..21Y,
2021ApJ...919...89Y}, and the nature of the central engine
\citep{2020ApJ...899L..27M, 2023MNRAS.520.1872B}. The current
consensus holds that magnetars are likely progenitors of FRBs
\citep{2020Natur.587...45Z, 2021SCPMA..6449501X,
2023RvMP...95c5005Z, 2023ApJS..269...17H}, largely because of the
observation of FRB 200428 from SGR 1935. As the only known
Galactic FRB source (noting that a recent study has reported
narrowband radio bursts from another Galactic magnetar, 1E
1547.0$-$5408, which appear to represent a low-energy analogue of
the repeating FRBs \citep{2026arXiv260321450L}) and the closest
one to us, SGR 1935 offers a unique laboratory for further
investigations, both in high-energy and in radio bands. Since
2020, high-energy monitoring campaigns have continued, leading to
the detection of additional X-ray bursts associated with radio
bursts \citep{2022ATel15690....1E, 2025ApJS..276...60R,
2025ApJS..277....5X, 2026ApJ...998L..44X, 2026A&A...707A.289T,
2026MNRAS.546ag312W} as well as independent X-ray bursts
\citep{2022ATel15674....1Y}. Likewise, radio monitoring across
multiple sub-bands has been ongoing, e.g., CHIME at 0.40--0.80 GHz
\citep{2022ATel15681....1D, 2022ATel15792....1P}, the Yunnan 40-m
radio telescope at 2.19--2.30 GHz \citep{2022ATel15707....1H}, and
the Robert C. Byrd Green Bank Telescope (GBT) at 4.00--8.00 GHz
\citep{2022ATel15697....1M}.

Thus, radio burst detections from SGR 1935 have now covered most
of the frequency range where FRBs are typically found
($\sim$0.1--8 GHz). However, evidence at the very low-frequency
end ($\sim$0.1 GHz) remains lacking. The main instruments capable
of probing this band are the LOw Frequency ARray (LOFAR) and the
Large Phased Array (LPA). LOFAR monitoring at 110--190 MHz in the
approximately two months following the epoch of FRB 200428 (though
not continuously) resulted in nondetections
\citep{2020ATel13707....1B, 2021MNRAS.503.5367B}. Similarly,
long-term LPA monitoring at 109--111.5 MHz from 2021 to 2025 also
yielded nondetections \citep{2025ARep...69..971F}. Nevertheless,
different claims exist regarding the LPA data taken in 2019--2021:
\citet{2021BLPI...48..317F} and \citet{2022ARep...66...32R}
reported an FRB-like pulse in the data of 2 September 2020, and
claimed the detection of pulsar-like (periodic) emission after
folding the data of January -- February and October -- December
2020, whereas \citet{Brylyakova2022, 2023ARep...67..163B} argued
that the signal-to-noise ratio (S/N) of the claimed FRB-like pulse
had been overestimated, i.e., it is a false detection. To clarify
this problem and to search for other potential bursts, we have
reanalyzed the LPA data of SGR 1935 obtained during two active
periods: March -- May 2020 (covering the epoch of FRB 200428) and
September -- November 2020 (covering the pulsar phase of October
2020). We employ new search methods tailored to both FRB-like and
pulsar-like signals.

This paper is organized as follows. Section \ref{sec2:pulse search
approaches} describes the LPA data and the methods we used for
single-pulse (i.e., matched-filter) and periodic pulse (i.e.,
folding) searches. Section \ref{sec3:search results and
constraints on the low-frequency radio behavior of SGR 1935+2154}
presents the search results and the derived constraints on the
low-frequency radio emission from SGR 1935. Specifically, we
derive upper limits on the energy of pulsar-like pulses, and for
FRB-like pulses we obtain burst rate upper limits under different
assumed energy distribution models.


\section{Pulse Search Approaches}
\label{sec2:pulse search approaches}

The LPA of the Lebedev Physical Institute of the Russian Academy
of Sciences, located at the Pushchino Radio Astronomy Observatory
(\href{https://www.prao.ru/index.php}{https://www.prao.ru/index.php})
near Moscow, is the world's largest meter-wave radio array, with a
total coverage area of 72,000 $\rm m^2$ and an effective area of
47,000 $\rm m^2$ at the zenith. It is also known as the BSA (Big
Scanning Antenna, a direct translation from its Russian name). As
a meridian-transit telescope, it relies on Earth's rotation to
survey the sky, covering a declination range from $-9^{\circ}$ to
$+55^{\circ}$, yielding about 3.5 minutes of daily monitoring for
SGR 1935, thereby filling a gap in long-term, low-frequency
observations of this Galactic FRB source. The LPA consists of
16,384 dipoles and 96 beams, with a beam size of $0.5^{\circ}
\times 1^{\circ}$. Its observing band is centered at 110.25 MHz
with a total bandwidth of 2.5 MHz \citep{2022MNRAS.517.1112T}. LPA
data are recorded simultaneously in two modes, offering different
time and frequency resolutions: (i) a low-time-resolution mode
with 6 channels (each 415 kHz wide) and a time resolution of
$\sim$100 ms; and (ii) a high-time-resolution mode with 32
channels (each 78 kHz wide) and a time resolution of $\sim$12.5
ms.

\begin{figure}[hbp]
\gridline{\fig{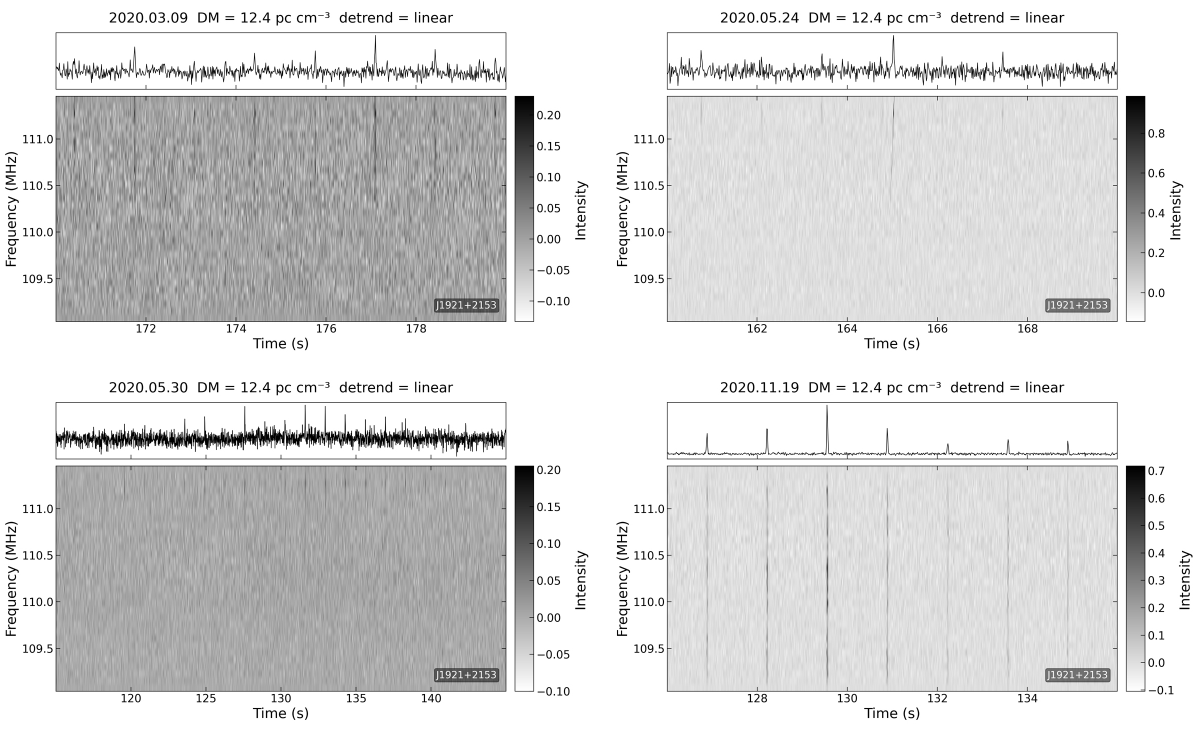}{1\textwidth}{}} \caption{
Contaminating pulses from PSR J1921+2153 observed through the LPA
side lobes and identified by the low-DM single-pulse pre-search.
The dynamic spectra on all four affected dates (9 March, 24 May,
30 May, and 19 November 2020) are shown. In the panel of 24 May,
the diagonal stripe covering the upper part of the LPA band at
around 165 s is radio frequency interference. The faint vertical
stripes above $\sim$111 MHz in that same panel, as well as the
signals in the other panels, correspond to genuine pulses from the
pulsar. } \label{figure1}
\end{figure}

As one of the few low-frequency survey arrays, the LPA has
provided valuable insights into our understanding of the
low-frequency radio behavior of pulsars
\citep{2011ARep...55..416K, 2015ATsir1628....1K,
2018RAA....18...97K, 2022MNRAS.517.1112T, 2024MNRAS.528.2220T},
rotating radio transients \citep{2022A&A...664A..37T,
2023ARep...67..590S}, and magnetars \citep{2005ARep...49..242M,
2006ChJAS...6b..68M}. In recent years, LPA survey data have also
been used to search for FRBs \citep{2018ATsir1641....1R,
2019ARep...63..877F, 2025PASA...42...59T}. The Galactic FRB source
SGR 1935, which is the focus of this paper, can produce both
FRB-like and pulsar-like pulses \citep{2023SciA....9F6198Z}.

We searched for both FRB-like single pulses and
pulsar-like periodic emission from SGR 1935 using the LPA
32-channel data taken during the two active epochs in 2020 (March
-- May and September -- November). Days heavily affected by
technical interruptions (routine LPA maintenance or power grid
failures), adverse weather (primarily lightning strikes), high
solar activity, and other radio frequency interference were first
discarded (about 16\% of all days). The interference level was
assessed by inspecting the noise track after calibration with the
standard calibration signal; given the moderate data volume, a
visual inspection of the discarded days was also performed. In
addition, the known pulsar J1921+2153 (also known as PSR B1919+21;
\citep{1968Natur.217..709H}), which has a DM of about 12.4 $\rm
pc~cm^{-3}$ and a period of about 1.34 s, sporadically enters the
LPA sidelobes. Its contaminating pulses were identified on four
days through a low-DM single-pulse pre-search (see Figure
\ref{figure1}) and subsequently masked in the original
time-frequency data. Because only the bright pulses of J1921+2153
are detectable via the LPA sidelobes, this cleaning step removes
all visible contamination without causing a significant loss of
exposure. Any residual pulses of J1921+2153 too weak to be
detected in the low-DM pre-search are buried in the noise, and
their large DM mismatch with SGR 1935+2154 (12.4 vs. 332.7 $\rm
pc~cm^{-3}$) causes them to be completely smeared out when the
data are dedispersed to the target DM, and therefore will not
affect the subsequent single-pulse and periodic-pulse searches.
Hereafter, we refer to this cleaned and masked data set as the
curated LPA data, which are the actual input for all subsequent
searches.

Because the features of the pulsar-like pulses
and the FRB-like pulses are quite different, we engage the
matched-filtering approach to search for FRB-like pulses, and use
folding approach to search for pulsar-like pulses, as detailed below.

\subsection{FRB-like Pulse Search via Matched Filtering}
\label{sec2.1:FRB-like pulse search via matched filtering}

Matched filtering is a standard method for searching for single pulses such
as FRBs \citep{2003ApJ...596.1142C}. To maximize our chances of detecting
faint single pulses, we first evaluate the broadening effects that
potential radio emission from SGR 1935 at $\sim$110 MHz may suffer, thereby
optimizing the matched-filter templates and tuning the search parameters to
improve detection efficiency and control the false-positive rate.

At low frequency, pulse broadening of radio waves is usually
significant. Specifically, the observed width of a dedispersed
single pulse can be described by \citep{2003ApJ...596.1142C,
2012hpa..book.....L, 2023RvMP...95c5005Z, 2025ApJ...980..114W}
\begin{eqnarray}
\label{eq1}
W_{\rm obs}=\sqrt{\left(1+z\right)^2 W_{\rm int}^2+\tau_{\rm sc}^2
+\tau_{\rm ins}^2},
\end{eqnarray}
where $W_{\rm obs}$ is the observed pulse width (after
dedispersion), $W_{\rm int}$ the intrinsic pulse width, $z$ the
redshift of the source (0 for SGR 1935), $\tau_{\rm sc}$ is the
temporal broadening due to plasma scattering (mainly from the
interstellar medium of the host galaxy and the Milky Way
\citep{2016ApJ...818...19K, 2023RvMP...95c5005Z}), and $\tau_{\rm
ins}$ is the instrumental broadening.

According to \citet{2016ApJ...832..199X}, for the Galactic source SGR 1935, $\tau_{\rm sc}$ can be estimated by
\begin{eqnarray}
\label{eq2}
\tau_{\rm sc}=0.2\left(\frac{\rm DM}{\rm 100~pc~ cm^{-3}}\right)^2
\left(\frac{\lambda}{\rm m}\right)^4~{\rm ms},
\end{eqnarray}
where $\lambda$ is the wavelength in the observer's frame. For the
LPA observations of SGR 1935, adopting a dispersion measure (DM)
of 332.7 $\rm pc~cm^{-3}$ \citep{2023SciA....9F6198Z,
2024ApJS..275...39W} and $\lambda=2.72~{\rm m}$ (the wavelength
corresponding to the LPA center frequency), we obtain $\tau_{\rm
sc}=121$ ms. When we take the wavelengths at the lower and upper
edges of the LPA operating band, the difference in scattering
broadening between the highest and lowest frequencies is
$\Delta\tau_{\rm sc}=11$ ms.

The instrumental broadening $\tau_{\rm ins}$ can be expressed as
\citep{2003ApJ...596.1142C, 2019A&ARv..27....4P, 2025ApJ...980..114W}
\begin{eqnarray}
\label{eq3}
\tau_{\rm ins}=\sqrt{\tau_{\rm DM}^2+\tau_{\rm \delta DM}^2
+\tau_{\Delta\nu}^2+\left(\Delta t\right)^2},
\end{eqnarray}
where $\tau_{\rm DM}$, $\tau_{\rm \delta DM}$, $\tau_{\Delta\nu}$ and
$\Delta t$ represent the frequency-dependent intra-channel dispersive
smearing, the residual dispersive smearing due to DM mismatch
\citep{2019ApJ...876L..23H}, the bandwidth smearing
\citep{1999ASPC..180..371B, 2018MNRAS.478.2337R}, and the sampling time of
the telescope, respectively. The first three terms can be computed by
\citep{2003ApJ...596.1142C}
\begin{eqnarray}
\label{eq4}
\tau_{\rm DM}=8.3\left(\frac{\rm DM}{\rm pc~cm^{-3}}\right)
\left(\frac{\Delta\nu}{\rm MHz}\right)
\left(\frac{\nu_{\rm c}}{\rm GHz}\right)^{-3}~{\rm \mu s},
\end{eqnarray}
\begin{eqnarray}
\label{eq5}
\tau_{\rm \delta DM}=\frac{\rm \delta DM}{\rm DM}\tau_{\rm DM},
\end{eqnarray}
\begin{eqnarray}
\label{eq6}
\tau_{\Delta\nu}=\left(\frac{\Delta\nu}{\rm MHz}\right)^{-1}~{\rm \mu s}.
\end{eqnarray}
Here $\Delta\nu$ is the channel bandwidth and $\nu_{\rm c}$ is the
central frequency of the channel. For the LPA observations of SGR
1935 in the 32-channel mode, $\Delta\nu = 2.5/32$ MHz, $\nu_{\rm
c}=110.25$ MHz, and we conservatively take ${\rm \delta DM}/{\rm
DM} = 0.01$ (overestimating the mismatch). This yields $\tau_{\rm
DM}=161$ ms, $\tau_{\rm \delta DM}=1.61$ ms, and
$\tau_{\Delta\nu}=0.01$ ms. Thus $\tau_{\rm DM}$ is the dominant
item among the instrumental contributions, while the other items
are all negligible. Similarly, using the lower and upper edges of
the LPA operating band in Equation (\ref{eq4}) gives a difference
in intra-channel dispersive smearing of $\Delta\tau_{\rm DM}=11$
ms across the band.

In summary, for FRBs which typically have intrinsic widths of a few to tens
of milliseconds \citep{2023ApJS..269...17H}, low-frequency observations not
only suffer from strong pulse broadening (i.e., $\tau_{\rm sc}$ and
$\tau_{\rm DM}$), but also exhibit a tail in the pulse profile due to the
frequency-dependent broadening differences (i.e., $\Delta\tau_{\rm sc}$ and
$\Delta\tau_{\rm DM}$). Consequently, such tails must be taken into account when
constructing matched-filter templates.

In fact, previous works involving LPA observations of SGR 1935
have already considered this issue. For example,
\citet{2021BLPI...48..317F}, \citet{2022ARep...66...32R} and
\citet{2023ARep...67..163B} used templates that incorporate
scattering-induced tails based on the empirical $\sim$110 MHz
scattering relation of \citet{2007ARep...51..615K}. However, their
implementations and underlying assumptions differ to some extent.

First, regarding the search strategy, \citet{2021BLPI...48..317F}
and \citet{2022ARep...66...32R} used templates with varying DM in
a blind search, while \citet{2023ARep...67..163B} employed a
fixed-DM template for a verification-oriented search. Second, the
data resolution is different: the former group used
low-time-resolution ($\sim$100 ms) six-channel data, whereas the
latter group used high-time-resolution ($\sim$12.5 ms) 32-channel
data. Third and most importantly, the two groups adopted different
assumptions about the intrinsic signal morphology at $\sim$110
MHz. \citet{2021BLPI...48..317F} and \citet{2022ARep...66...32R}
assumed that the signal itself is scattered to second-long widths
at 110 MHz so that they used second-long templates to recover the
full pulse energy. By contrast, \citet{2023ARep...67..163B}
argued, based on observations of SGR 1935 at other wavebands
\citep{2020Natur.587...54C, 2020Natur.587...59B,
2021NatAs...5..414K}, that the intrinsic pulse width is extremely
narrow (milliseconds), far smaller than the scattering time at
$\sim$110 MHz calculated through the Kuzmin scattering function.
In this case, the observed signal should be an impulse response of
a narrow pulse convolved with the Kuzmin scattering function, and
its effective duration is short. In their viewpoint, using
second-long templates would inevitably introduce excessive noise
outside the signal window, degrading rather than enhancing S/N. To
reinforce this point, \citet{2023ARep...67..163B} performed a
limit test: assuming the ``scattered to seconds'' hypothesis, they
deliberately constructed a wider template using DM = 375 $\rm
pc~cm^{-3}$ (higher than the true DM = 332.7 $\rm pc~cm^{-3}$) and
searched the high-resolution data. Even with this wider template,
no signal was found, firmly excluding the possibility that a
template mismatch (too narrow) could lead to nondetection.

\begin{figure}[htbp]
\gridline{\fig{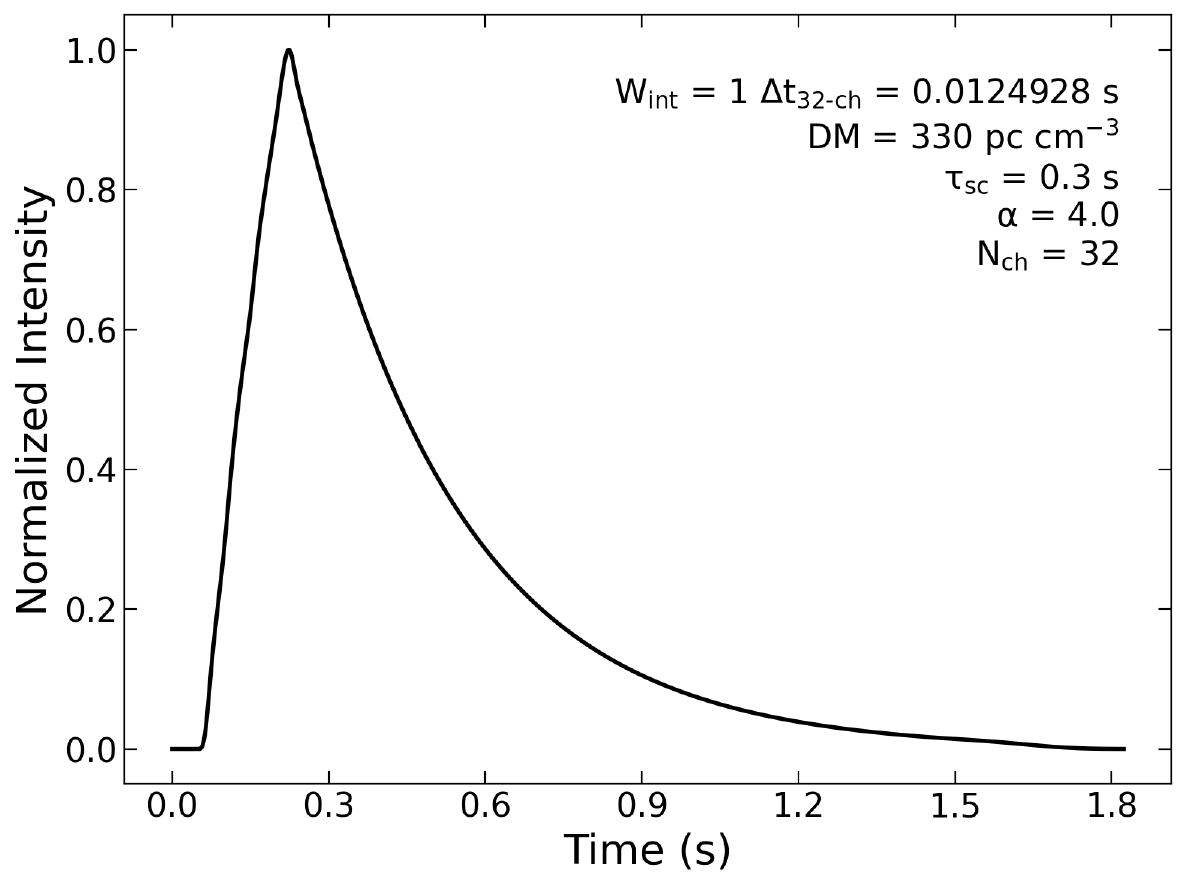}{0.6\textwidth}{}} \caption{ Example of
a matched filter template generated by our customised method. The
template is produced using a specific set of parameters (intrinsic
pulse width $W_{\rm int}$, DM, scattering time constant $\tau_{\rm
sc}$, and scattering index $\alpha$) for the 32-channel data
($N_{\rm ch}=32$) of LPA. It illustrates the typical morphology
expected for a dedispersed scattered pulse at LPA frequencies. In
practice, our template library contains many different
combinations of the parameters, aiming to search for FRB-like
pulses over a broad parameter space. } \label{figure2}
\end{figure}

Nevertheless, it must be noted that SGR 1935 has no prior
low-frequency radio detections, so we lack any reliable knowledge
of the pulse width at $\sim$110 MHz. Using observations at other
radio bands to infer its low-frequency pulse morphology is of
limited usage. In light of this, we develop a more generic,
customized search method. The core idea is not to assume any
particular intrinsic width $W_{\rm int}$ or fixed scattering
relation, but to simultaneously explore a multi-dimensional
parameter space that includes the intrinsic pulse width $W_{\rm
int}$ (ten logarithmically spaced values from milliseconds to
seconds), the scattering time constant $\tau_{\rm sc}$ (six
independent values: 0.01, 0.03, 0.1, 0.3, 1.0, 2.0 s), the
scattering frequency index $\alpha$ (five values: 3.5, 3.75, 4.0,
4.25, 4.5), and the DM (scanned around the known value of 332.7
$\rm pc~cm^{-3}$, within $\pm 20~{\rm pc~cm^{-3}}$ in steps of 0.5
$\rm pc~cm^{-3}$). This produces a comprehensive template library
for a grid search. Unlike previous works that relied on a fixed
empirical $\sim$110 MHz scattering relation $\tau_{\rm
sc}=0.06\left(\frac{\rm DM}{\rm 100~pc~cm^{-3}}\right)^{2.2} ~{\rm
s}$ (which ties $\tau_{\rm sc}$ to DM and implicitly assumes
$\alpha\approx4$ \citep{2007ARep...51..615K}), we allow $\tau_{\rm
sc}$ and $\alpha$ to vary independently. Following the standard
thin-screen scattering model \citep{2009tra..book.....W}, the
frequency-dependent scattering time for a channel with central
frequency $\nu_{\rm c}$ is
\begin{eqnarray}
\label{eq7}
\tau_{\nu_{\rm c}}=\tau_{\rm sc}\left(\frac{\nu_{\rm c}}
{{\mathcal{V}}_{\rm c}}\right)^{-\alpha},
\end{eqnarray}
where ${\mathcal{V}}_{\rm c}=110.25$ MHz is the center frequency
of LPA. Following previous studies, we adopt a one-sided
exponential decay as the scattering convolution kernel:
\begin{eqnarray}
\label{eq8}
k\left(t,\nu_{\rm c}\right)=\frac{1}{\tau_{\nu_{\rm c}}}
{\rm e}^{-t/\tau_{\nu_{\rm c}}},\quad t\geq0.
\end{eqnarray}

For each parameter set, we generate a simulated dynamic spectrum,
apply incoherent dedispersion, and average over channels to
produce a one-dimensional time series that serves as the template.
Figure \ref{figure2} shows an example of the template generated
under a specific set of parameters. The key differences between
our method and previous studies are:\

(1) Previous studies determined $\tau_{\rm sc}$ from a
DM-dependent scattering relation and directly substituted it for
$\tau_{\nu_{\rm c}}$ in Equation (\ref{eq8}) to obtain the
convolution kernel, which was then convolved with a square-wave
intrinsic pulse to generate a one-dimensional template.\

(2) We take both $\tau_{\rm sc}$ and $\alpha$ as free parameters,
and use Equation (\ref{eq7}) to compute $\tau_{\nu_{\rm c}}$ of
each channel. A square-wave intrinsic pulse (whose width $W_{\rm
int}$ is also a free parameter) is then convolved with the
channel-dependent kernel to produce a simulated dynamic spectrum,
which is dedispersed and averaged over channels to obtain the
final template.

Thus, our approach transforms the prior question of ``what shape should the
signal have?'' into a posterior choice determined by the data themselves,
minimizing reliance on uncertain assumptions about pulse width and
scattering relations, and allowing a more data-driven evaluation of
candidate signals.

\begin{figure}[htbp]
\gridline{\fig{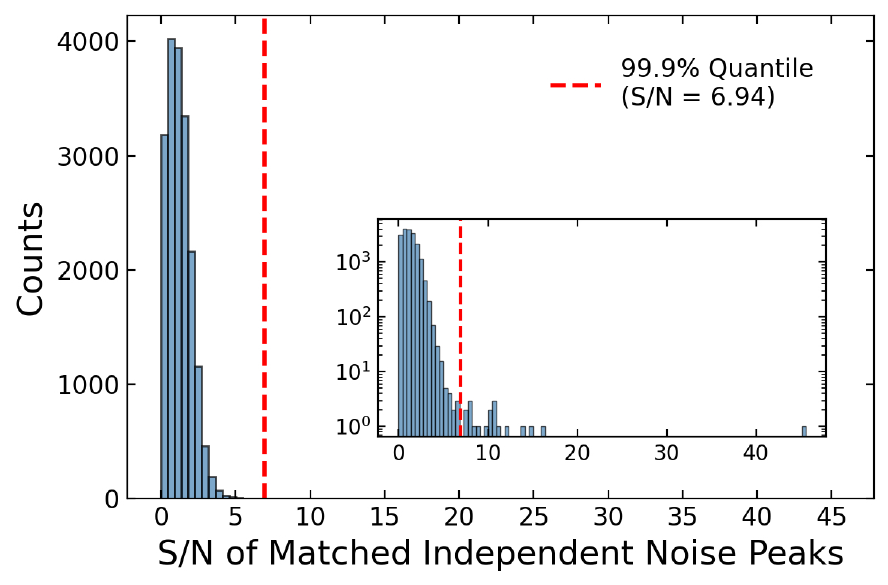}{0.6\textwidth}{}}
 \caption{ Post-trial S/N distribution of independent noise peaks
obtained from five days of pure-noise LPA data, after dedispersion at
DM = 0 and processing with our full matched-filter template bank. The
vertical dashed line marks the 99.9th percentile (S/N = 6.94), which we
conservatively round up to S/N = 7 as the detection threshold for our
single-pulse search. The inset displays the same distribution on a
logarithmic vertical scale to highlight the high-S/N tail. }
\label{figure3}
\end{figure}

To validate our customized single-pulse search method, we
performed a post-trial false-alarm analysis and an injection-based
recovery-fraction test. For the post-trial false-alarm analysis,
we constructed a pure-noise base by selecting five days of clean,
moderate-quality observations from the curated LPA data and
dedispersing them at DM = 0. This pure-noise base contains no
astrophysical signal and share the same noise statistics as the
actual search data. We then processed it with exactly the same
matched-filter template bank used in the actual search, covering
all parameter combinations of intrinsic pulse width, scattering
time constant, scattering frequency index, and DM, and recorded
all single-pulse candidates (i.e., noise peaks) and their S/N
values. Because a single noise peak can produce multiple
candidates at adjacent times and through different templates, we
applied an adaptive time-proximity clustering algorithm, in which
the merging window for any two candidates is set proportional to
the larger equivalent width of their respective detection
templates, and each cluster retains only the highest-S/N candidate
as an independent noise peak event. The distribution of the S/N
values for all independent noise peaks is presented in Figure
\ref{figure3}, reflecting the statistical behavior of the noise
peaks in our curated LPA data. The 99.9th percentile of this
distribution is S/N = 6.94; we therefore adopted S/N = 7 as the
detection threshold, giving a probability of less than 0.1\% that
a single-pulse candidate with S/N $>$ 7 arises from noise
fluctuations alone.

\begin{figure}[htbp]
\gridline{\fig{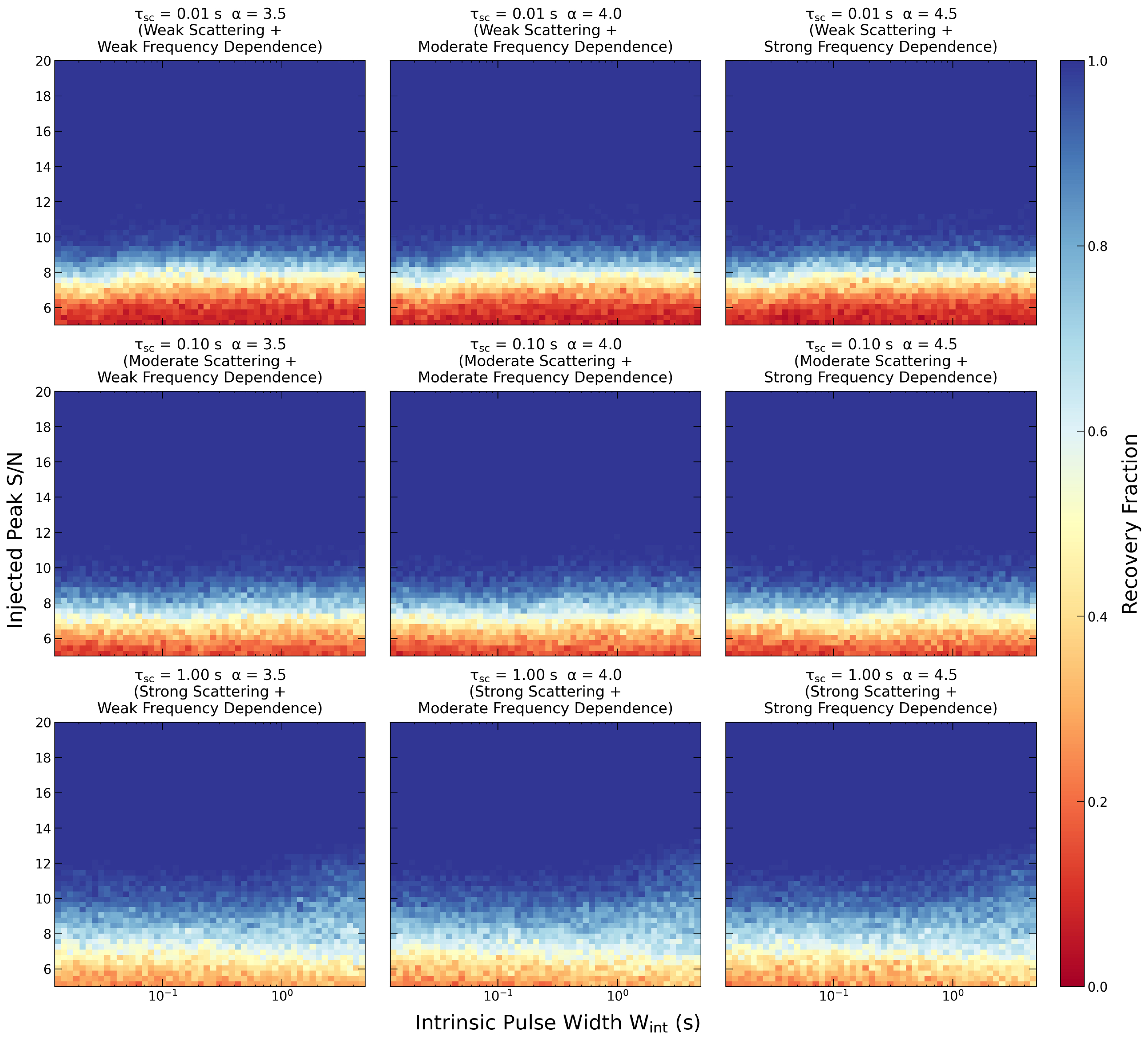}{1\textwidth}{}}
 \caption{ Recovery-fraction heatmaps from the artificial-injection
test. The injection base consisted of a single daily observation taken
from the LPA data, chosen for its root-mean-square noise level close to
the median of the full data set, and was dedispersed to the DM of SGR
1935+2154 (332.7 $\rm pc~cm^{-3}$) prior to the artificial pulse
injection. The nine panels cover the combinations of weak, moderate, and
strong scattering with weak, moderate, and strong scattering frequency
dependence, each corresponding to a fixed pair of scattering time
constant ($\tau_{\rm sc}$) and scattering frequency index ($\alpha$).
Within each panel the recovery fraction is plotted as a function of
intrinsic pulse width ($W_{\rm int}$) and injected peak S/N; each pixel
represents 100 independent trials, where a single artificial pulse is
injected per trial. The color bar indicates the values of
the recovery fraction. }
\label{figure4}
\end{figure}

\begin{figure}[htbp]
\gridline{\fig{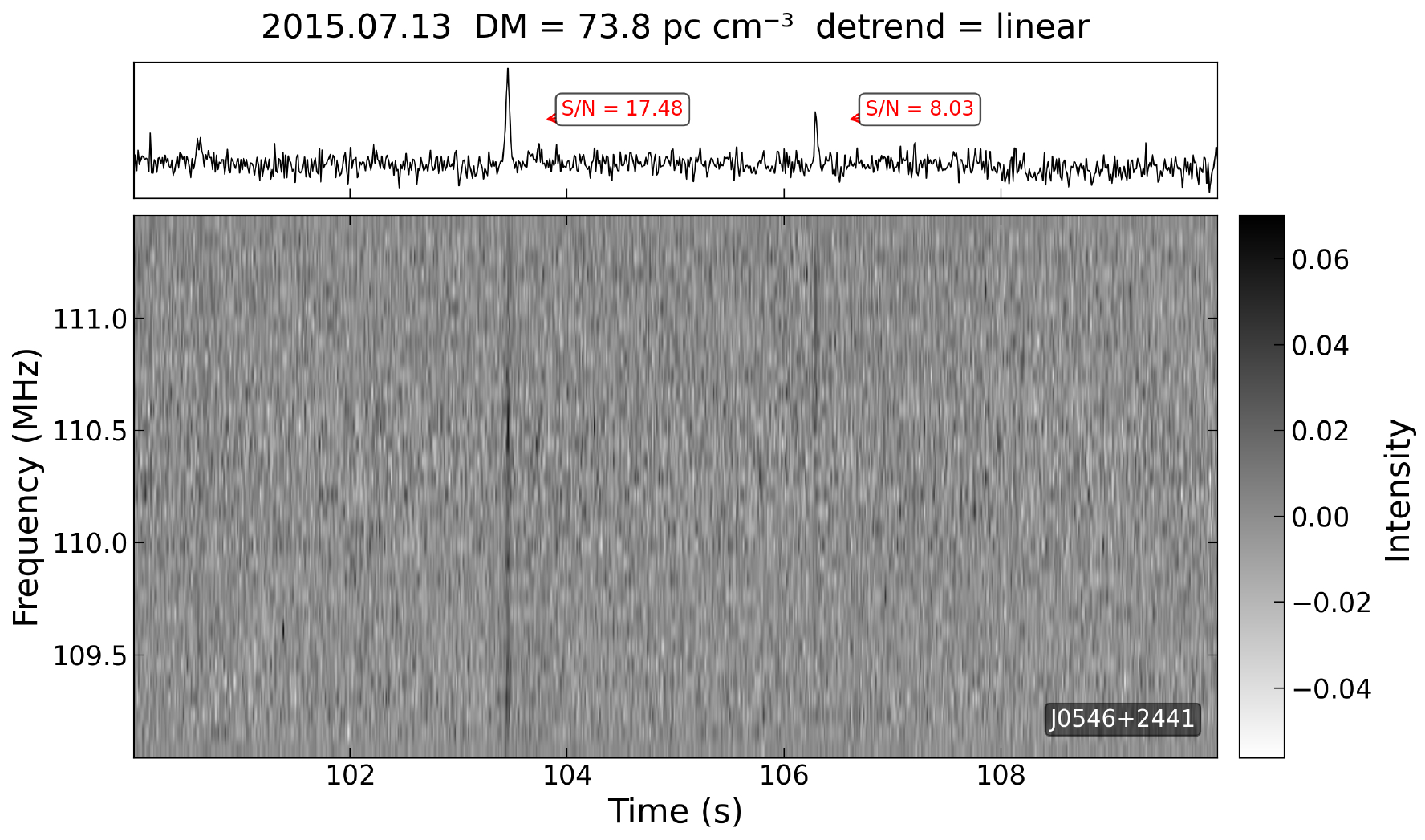}{1\textwidth}{}}
 \caption{ Validation
of our new single-pulse search method using pulsar J0546+2441. Two
individual pulses were revealed in LPA 32-channel data taken on 13
July 2015 (UTC time, same hereafter), with signal-to-noise ratios
(S/N) of 17.48 and 8.03, respectively. The detection demonstrates
that our method works well for pulses either with or without
significant trailing features. }
\label{figure5}
\end{figure}

For the recovery fraction test based on artificial injection, we
selected a single daily observation segment from the curated LPA data,
whose root-mean-square noise level is close to the median of the
full data set, to serve as the injection base. We dedispersed it at the
DM of SGR 1935 (332.7 $\rm pc~cm^{-3}$) and then applied exactly the
same preprocessing as in the actual search, including detrending,
bad-block removal, and normalizing. Then, artificial pulses were
generated using the same physical model that was employed to construct
the matched-filter template bank and injected at random time locations. For each
injected pulse, we ran the full single-pulse search pipeline and checked
whether a candidate was found within a narrow time window around the
injection time location with S/N $>$ 7; if so, the pulse was considered
recovered.

To probe the template bank's response to diverse pulse
morphologies, we scanned over the intrinsic pulse width, the
injected peak S/N (the most direct detection statistic in our
pipeline and the variable that avoids flux calibration
uncertainties), the scattering time constant, and the scattering
frequency index. The resulting recovery fractions of the scanned
parameter combinations are summarized in Figure \ref{figure4},
which displays nine heatmaps, each corresponding to a fixed pair
of scattering time constant ($\tau_{\rm sc}$) and scattering
frequency index ($\alpha$). These nine panels cover the
combinations of weak, moderate, and strong scattering with weak,
moderate, and strong scattering frequency dependence. Within each
panel, the recovery fraction is shown as a function of the
intrinsic pulse width ($W_{\rm int}$) and the injected peak S/N;
each pixel represents the recovery fraction after 100 independent
trials, with the value encoded by the color bar. Across all
panels, the recovery fraction shows little dependence on $W_{\rm
int}$, as expected because our template bank already spans a wide
range of intrinsic pulse widths, and rises steadily with
increasing injected peak S/N, climbing from $\sim$0.5 to 1 above
S/N $\approx$ 7 and falling off quickly below that. The recovery
fraction is largely insensitive to $\alpha$ but depends measurably
on $\tau_{\rm sc}$, such that a slightly higher injected peak S/N
is required to reach a recovery fraction of exactly 1 for the
largest $\tau_{\rm sc}$ values; nevertheless, the overall recovery
fraction for injected peak S/N $>$ 7 remains high. The occasional
detection of injected pulses below the S/N = 7 threshold occurs
because the local noise at the injection time location can be
lower than the globally averaged noise used to compute the
injected peak S/N. Such injection test demonstrates that our
single-pulse search method is robust and trustworthy, and is
sensitive to a broad range of pulse morphologies.

We have also tested our search method on real LPA observations of known
pulsars \citep{2024MNRAS.528.2220T}. For instance, our method reliably
recognizes individual pulses from PSR J0546+2441 which was originally
discovered by \citet{2005MNRAS.363..929C}. Figure \ref{figure5}
presents two pulses (including a faint one) recognized by using
our method in LPA 32-channel data of J0546+2441 taken on 13 July
2015, with S/N values of 17.48 and 8.03, respectively. While the
injection test has quantitatively demonstrated the sensitivity of our
method to diverse pulse morphologies, Figure \ref{figure5} provides an
illustrative example of its application to real LPA data.

\subsection{Pulsar-like Pulse Search via Folding}
\label{sec2.2:pulsar-like pulse search via folding}

\subsubsection{Periodicity Search}
\label{sec2.2.1:periodicity search}

We use period folding method to search for pulsar-like pulse from
SGR 1935. Before performing the folding, we first need to
determine the candidate periods. SGR 1935 is not a canonical
pulsar, although it occasionally exhibits pulsar-like emission
\citep{2023SciA....9F6198Z, 2024ApJS..275...39W}. Our data cover a
nine-month baseline (March -- May and September -- November 2020),
which will inevitably lead to difficulties due to phase-connection
and error-accumulation issues. This is especially true given that
SGR 1935 was in an active phase between March and November 2020,
producing several short-timescale spin changes (e.g., three spin
changes in ten days in October 2020 \citep{2023NatAs...7..339Y,
2024Natur.626..500H}) and making a joint period search extremely
challenging. On the other hand, periodicity search based on
single-day data analysis would result in low significance for any
candidate periods due to the daily $\sim$3.5-minute observation
segments. As a compromise, we adopt the following approach: we
perform an independent period search on each daily segment
(ignoring the period derivative $\dot{P}$ because of the short
time span) and apply a significance threshold of $5\sigma$. After
collecting all candidates with significance $>5\sigma$ from all
days, we arrange them into a one-dimensional array and compute the
probability density distribution. For each DM value used in
dedispersion, we obtain one such distribution. By comparing these
distributions across different DMs, regions where multiple
distributions overlap at high probability densities indicate
candidate period ranges. Within those ranges, we then identify the
period values that peak at different DM curves simultaneously as
our final candidate periods (i.e., the most probable periods).

It should be mentioned that although the period $P$, period
derivative $\dot{P}$, and reference MJD of SGR 1935  during its
October 2020 pulsar-like activity have been reported previously
\citep{2023SciA....9F6198Z}, we take a fully blind search on each
daily segment. This is because our data have a nine-month
baseline. Also, SGR 1935 is an active magnetar likely exhibiting
complex rotational irregularities and we have no prior knowledge
of its low-frequency behavior during the 2020 active episodes.
This strategy aligns with the data-driven principle described
above. It also motivates the relatively wide DM window
($332.7\pm20~{\rm pc~cm^{-3}}$) used in the period search. Unlike
single-pulse searches, period searches are less sensitive to
slight DM mismatches that cause pulse broadening. Thus, our use of
a deliberately wide window guarantees that any real and
significant periodic signal, if present, remains detectable even
under a tolerable level of DM mismatch.

In addition, we employ two different period-search methods, the
Fast Fourier Transform (FFT) and the Fast Folding Algorithm (FFA).
FFT is computationally efficient but suffers from limited
frequency resolution (due to the short duration of each segment)
and is prone to red-noise contamination when searching for
low-frequency periods; hence, it is suitable for a quick
preliminary scan. FFA, in contrast, is not restricted to the
discrete Fourier frequency bins and is generally less susceptible to red
noise than FFT, making it more sensitive to long-period signals and
complex pulse shapes. Although its achievable period resolution and
sensitivity are still fundamentally limited by the finite observing
span, and it is not completely immune to red noise, nor to other sources
of interference such as radio frequency interference, its complementary
strengths make it well suited for verifying candidates identified by FFT
and for detecting signals that FFT may have missed.

For the FFT-based search, we use the \texttt{accelsearch}
routine from the \texttt{PRESTO} package \citep{2011ascl.soft07017R}.
Each 3.5-minute daily time series, with a sampling time of 12.5 ms,
consists of approximately 16,500 samples. A harmonic summing of up to 8
harmonics is applied, which enhances sensitivity to narrow pulse
profiles (down to a duty cycle of $\sim$0.01). No acceleration search is
performed, as the short integration time renders the effect of the
spin-period derivative negligible. For the FFA-based search, we fold each
daily time series over a trial period grid spanning 0.1 to 8.0 s with a
step size of 0.01 s (comparable to the time resolution of our 
data). The folded profiles are accumulated into 64 phase
bins, which naturally accommodates pulse widths up to a duty cycle of
$\sim$0.2, and the significance of
each profile is evaluated using its S/N. From both FFT and FFA searches,
the most probable periods are determined by identifying where the
probability density distributions of candidate periods from different DM
trials overlap, as described earlier, yielding statistical period
estimates whose precision can far exceed the original FFT/FFA grid
resolution, being instead governed by the number of candidate periods.
Combining FFT and FFA searches, we cover duty cycles ranging from narrow
pulses ($\sim$0.01) to broad profiles ($\sim$0.2), where FFT 
handles the narrower end and FFA handles the broader end. No 
downsampling is applied to the time series prior to the period searches.

\subsubsection{CCF-assisted Folding Method}
\label{sec2.2.2:CCF-assisted folding method}

Once the most probable periods are obtained, the next step is pulse
folding. For reasons already discussed (nine-month baseline, SGR
1935 being an active and unstable magnetar, unknown low-frequency
behavior), joint folding of all daily segments, even with
$\dot{P}$ accounted for, is impractical. Therefore, we adopt an
empirical phase-concatenation method. We fold each daily
3.5-minute segment independently (ignoring $\dot{P}$ because of
the short duration), then compute the cross-correlation function
(CCF) between each pair of daily folded profiles to determine the
optimal phase shift, and finally shift and stack the profiles to
maximize any potential pulse feature. The advantage of this
approach is that it can detect weak periodic signals. The
disadvantage is that it may produce false signals.

\begin{figure}[htbp]
\gridline{\fig{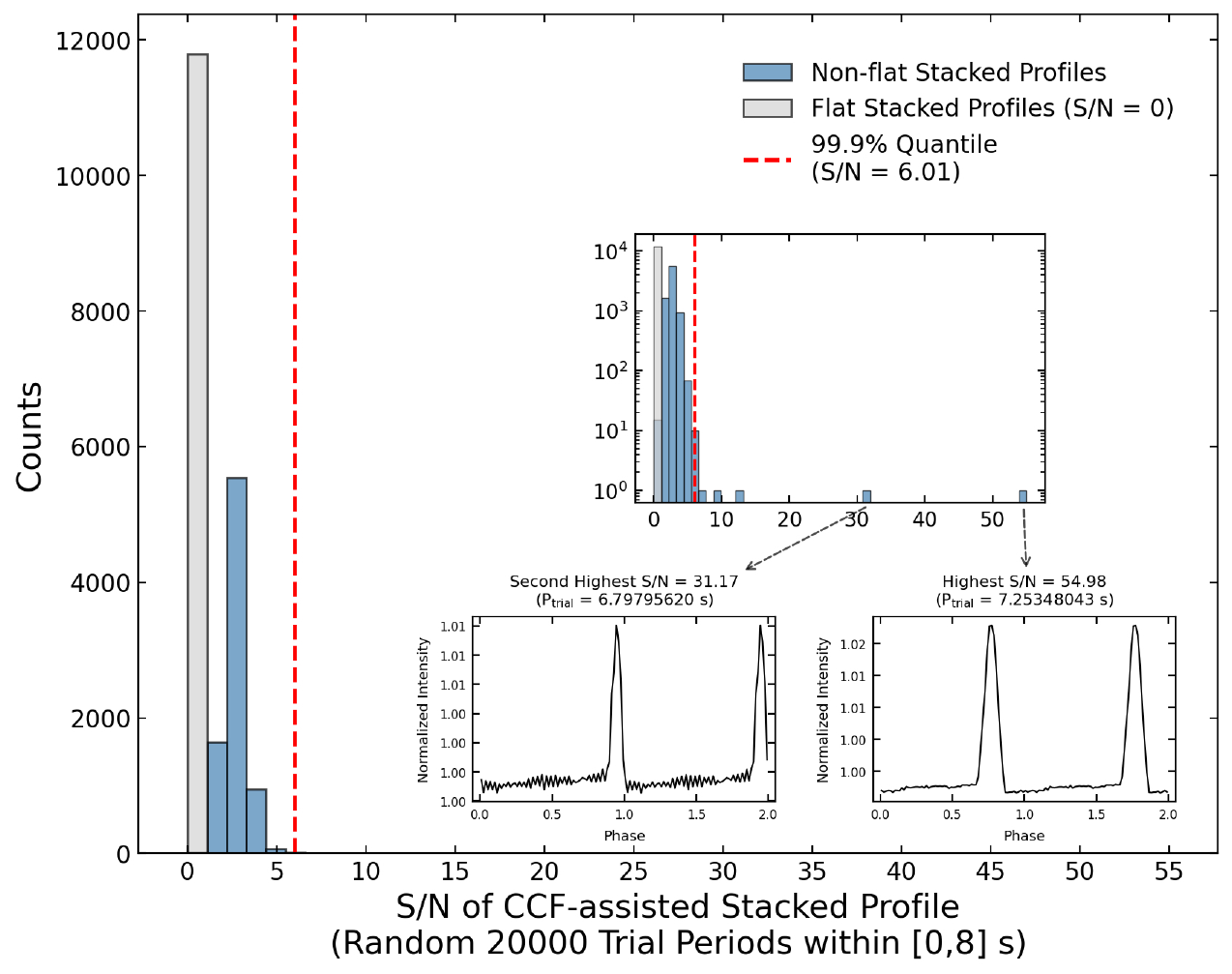}{0.6\textwidth}{}}
 \caption{ Chance coincidence analysis of the CCF-assisted folding
method. The distribution of S/N values is obtained from 20,000 random
trial periods, drawn from a uniform distribution on
$\left[0,8\right]$ s, applied to the zero-DM pure-noise base covering
all available days of the curated LPA data. The gray histogram shows
flat stacked profiles (kurtosis $<-0.5$, assigned S/N = 0), while the
blue histogram shows non-flat profiles that passed the kurtosis test.
The vertical dashed line marks the 99.9\% quantile of the S/N
distribution for non-flat profiles (S/N = 6.01), which we adopt as the
reliability threshold (S/N = 6) for CCF-assisted stacking, corresponding
to a chance coincidence probability of 0.1\%. The upper inset displays
the same distribution as in the main plot but on a logarithmic vertical
scale to highlight the high-S/N tail. The lower insets display the two
highest-S/N stacked profiles. }
\label{figure6}
\end{figure}

\begin{figure}[htbp]
\gridline{\fig{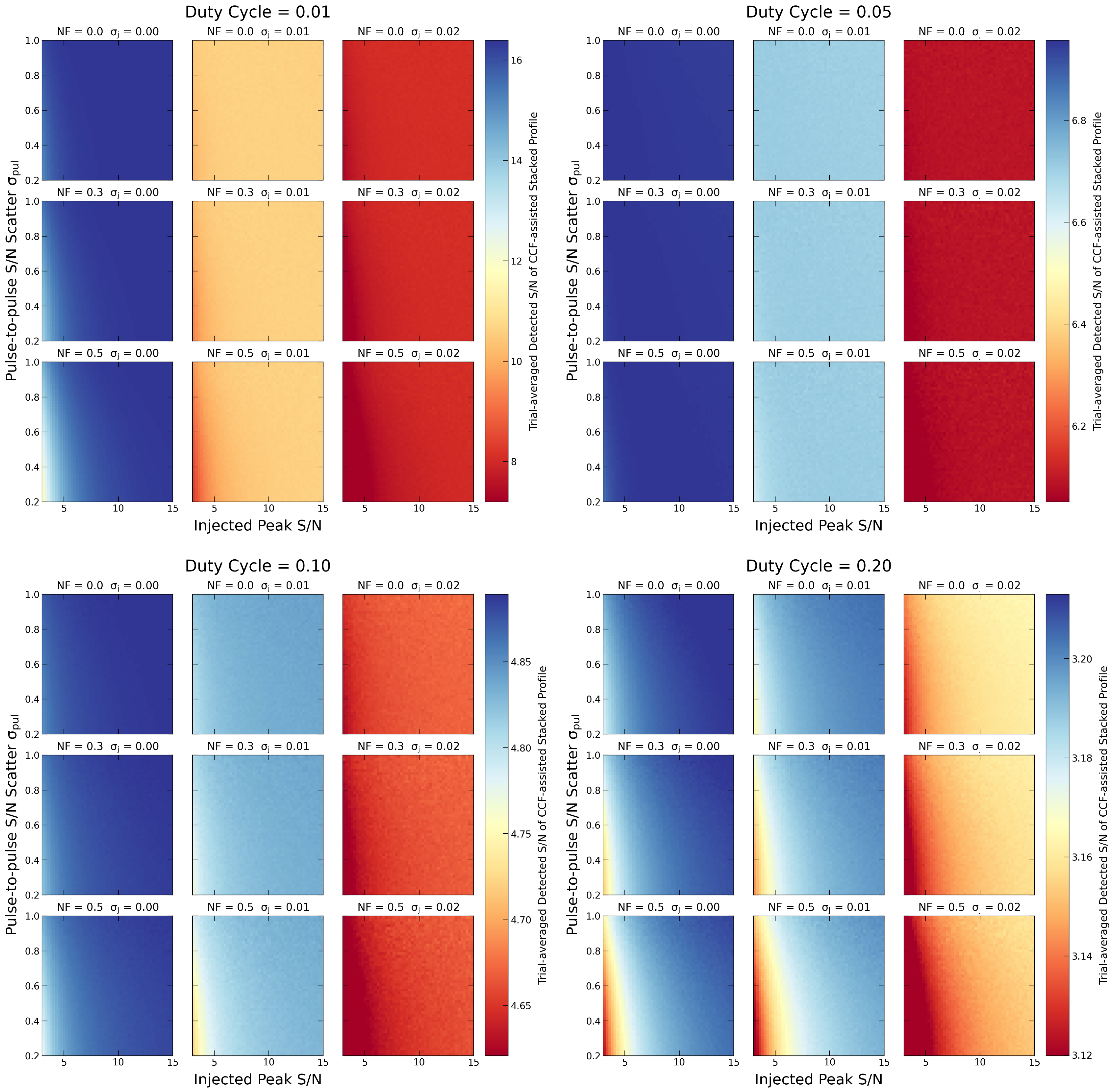}{1\textwidth}{}}
 \caption{ Results of the periodic signal injection test. The four major
blocks correspond to different duty cycles (0.01, 0.05, 0.10, 0.20).
Within each block, the nine heatmaps form a $3\times3$ grid covering the
combinations of nulling fraction (NF = 0, 0.3, 0.5) and phase jitter
($\sigma_{\rm j} = 0, 0.01, 0.02$). In each heatmap, the horizontal axis
gives the reference injected peak S/N and the vertical axis shows the
pulse-to-pulse S/N scatter $\sigma_{\rm pul}$. For each individual
pulse, its actual peak S/N is the reference value multiplied by a
log-normal random factor (median 1, logarithmic standard deviation
$\sigma_{\rm pul}$) and its phase is perturbed by a Gaussian offset
$\mathcal{N}(0,\sigma_{\rm j})$. Each pixel represents the mean detected
S/N of the CCF-assisted stacked profile averaged over 100 independent
trials, as indicated by the color bar. }
\label{figure7}
\end{figure}

To validate this CCF-assisted folding method, we also performed
a chance coincidence analysis and an injection test with simulated
periodic signals. For the chance coincidence analysis, to fully capture
the noise behavior of the CCF-assisted stacking over the nine-month
baseline, we constructed a pure-noise base using all available days of
the curated LPA data after dedispersion at DM = 0. Because this DM value
is far from the true DM of SGR 1935 (332.7 $\rm pc~cm^{-3}$) and the
known pulsar contamination has been masked, this noise base can be
regarded as containing no real astrophysical signal. For each of 20,000
random trial periods drawn from a uniform distribution on
$\left[0,8\right]$ s, we performed CCF-assisted folding on the noise
base across all days. The resulting stacked profiles were first
subjected to a kurtosis test: if the profile kurtosis was below $-0.5$,
indicative of a flat, featureless shape, we assigned it an S/N of zero;
only profiles that passed this test were then evaluated by computing the
S/N of the peak profile. The obtained distribution of S/N values is
shown in Figure \ref{figure6}. The vast majority of trials produce flat
stacked profiles (gray histogram), demonstrating that the CCF-assisted
folding method does not systematically inflate the significance of noise
peaks. For the S/N distribution of the non-flat stacked profiles (blue
histogram), the 99.9th percentile is 6.01. We therefore adopt S/N = 6 as
the reliability threshold for CCF-assisted stacked profiles, where the
chance coincidence probability that a noise-only profile passing the
kurtosis test exceeds this threshold is 0.1\%. We note that a small
number of spurious stacked profiles appear in the high-S/N tail of the
distribution (see insets of Figure \ref{figure6}).  Such occurrences are
statistically expected and are explicitly accounted for by the 0.1\%
chance coincidence probability associated with our S/N = 6 threshold.
Hence, these rare events do not compromise the reliability of our
detection criterion.

For the injection test, we injected simulated pulsar-like
signals into the zero-DM noise base covering all available days of the
curated LPA data set, the same as employed in the chance coincidence
analysis, adopting the known spin period of SGR 1935 (3.24781628 s, as
reported by \citealt{2023SciA....9F6198Z}). To reflect the diversity of
real pulsar-like signals, we scanned a broad parameter space designed to
capture their typical observed properties, including both static
morphology and pulse-to-pulse variations. For each parameter
combination, we conducted 100 independent trials. In each trial, a
randomly generated series of periodic pulses (matching the given
parameter set) was injected into the noise base; we then ran the full
CCF-assisted folding pipeline on the simulated data and measured the S/N
of the resulting stacked profile. Over the 100 trials, we averaged the
measured S/N values to obtain the mean detected S/N for that parameter combination.

The results of the periodic signal injection tests are
summarized in Figure \ref{figure7}. We see that the detected S/N of the CCF-assisted
stacked profile decreases systematically with increasing duty cycle, and
it remains consistently above the reliability threshold of S/N = 6 for
duty cycles below 0.1. The nulling fraction has a negligible effect,
whereas stronger phase jitter leads to a systematic decline in the
recovered S/N. Within each heatmap, neither the pulse-to-pulse S/N
scatter nor the reference injected peak S/N significantly affects the
detected S/N; the only notable exception is that, at low reference
injected peak S/N, a larger $\sigma_{\rm{pul}}$ tends to raise the
detected S/N, because it occasionally produces some individual pulses
with higher peak S/N. These results show that our method reliably
recovers periodic signals across a broad range of morphologies, with
detection efficiency primarily limited by the duty cycle and phase jitter.


In practice, to further guard against the influence of small
random noise fluctuations on the final stacked profile, we applied a
kurtosis-based filter to implement a branched folding strategy. Daily
profiles with kurtosis below $-0.5$ were classified as flat. Only when
the number of days with non-flat profiles exceeded 50 (roughly one-third
of the days of the curated LPA data) did we perform CCF-assisted folding
on all daily profiles; otherwise, we adopted a conservative, standard
$\dot{P}$-corrected folding procedure on the global time series over all
days, using the $\dot{P}$ value of SGR 1935 ($3.717\times10^{-11}~{\rm{s}}~
{\rm{s^{-1}}}$) reported by \citet{2023SciA....9F6198Z}. Importantly, if
the CCF-assisted folding, namely a phenomenological stacking, yields a
clear pulse peak, its S/N should be interpreted as an upper limit (i.e.,
the true S/N could be lower), because the alignment procedure
deliberately boosts the peak. If no peak emerges, it rules out 
the presence of pulsar-like pulses above our search sensitivity and 
within the explored parameter space.


\clearpage

\section{Search Results and Constraints on SGR 1935+2154}
\label{sec3:search results and constraints on the low-frequency radio
behavior of SGR 1935+2154}

\subsection{FRB-like Pulses}
\label{sec3.1:FRB-like pulses}

For the curated LPA 32-channel data of SGR 1935 during March -- May and
September -- November 2020, we performed single-pulse search using our
customized matched-filtering method. In practice, when a candidate
burst exceeded our detection threshold (S/N $>$ 7), we generated a
dedispersed dynamic spectrum based on the corresponding time range
and trial DM, and visually inspected it to examine the presence of
a genuine burst. No significant burst is detected from SGR 1935 in
our analysis.

This nondetection, combined with the LPA sensitivity to single pulses and
Monte Carlo (MC) simulations, allows us to place model-dependent upper
limits on the burst rate of FRB-like pulses from SGR 1935 at $\sim$110
MHz. The LPA single-pulse sensitivity, $\mathcal{S}_{\rm sp}$, is
calculated following \citet{2022MNRAS.517.1112T} and \citet{ 2021ApJ...923..230L} as
\begin{eqnarray}
\label{eq9}
\mathcal{S}_{\rm sp}=\frac{\left({\rm S/N}\right)_{\rm min,sp}T_{\rm sys}}
{G\sqrt{N_{\rm pol}\Delta\mathcal{V}W_{\rm obs}}},
\end{eqnarray}
where $\left({\rm S/N}\right)_{\rm min,sp}=7$ is the adopted
single-pulse detection threshold, $T_{\rm sys}=1000$ K is the system
temperature (the sky temperature at 110 MHz is stable for a fixed
direction, and ambient-induced variations of the receiver temperature
are within $\pm$2\%, which we consider negligible),
$G=17~{\rm K~{Jy}^{-1}}$ is the gain, $N_{\rm pol}=1$ is the number of
polarizations, $\Delta\mathcal{V}=2.5$ MHz is the operating bandwidth, and
$W_{\rm obs}$ is the observed pulse width. Equation (\ref{eq9}) shows that a
larger $W_{\rm obs}$ yields a smaller $\mathcal{S}_{\rm sp}$ (i.e., a
higher sensitivity). To obtain a conservative estimate of the upper limit
on burst rate, we adopt the smallest observable $W_{\rm obs}$ set by the
LPA time resolution in 32-channel mode, i.e., 12.5 ms. This maximizes
$\mathcal{S}_{\rm sp}$ (makes it least sensitive) and therefore leads to
the most conservative burst-rate upper limits. Using a more sensitive
threshold would yield artificially lower (hence less conservative) limits.
With these values, Equation (\ref{eq9}) gives $\mathcal{S}_{\rm sp}=2.33$ Jy.

We then estimate the burst-rate upper limits for FRB-like pulses
by assuming two energy-distribution models, including a log-normal
distribution \citep{2023ApJS..269...17H} and a power-law
distribution. First, we extract and merge all valid time windows
from the data to build the total coverage interval. For a given
set of distribution parameters $\theta$ (for the log-normal model:
$f\left(S_\nu\right)=\frac{1}{\sqrt{2{\rm \pi}} \sigma
S_\nu}\exp{\left[-\frac{\left(\ln{S_\nu}-\mu\right)^2}
{2\sigma^2}\right]}$, $S_\nu>0$, $\theta=\left(\mu,\sigma\right)$;
for the power-law model:
$f\left(S_\nu\right)=\frac{\beta-1}{S_{\nu{\rm ,min}}}
\left(\frac{S_\nu}{S_{\nu{\rm ,min}}}\right)^{-\beta}$,
$S_\nu>S_{\nu{\rm ,min}}$, $\theta=\left(S_{\nu{\rm
,min}},\beta\right)$), we perform $N=5000$ MC simulations for each
burst rate $\Lambda$ on a logarithmic grid. In each simulation, we
generate event times following a Poisson process (with
inter-arrival times $h$ following an exponential distribution,
i.e., $f\left(h\right)=\Lambda\exp{\left(-\Lambda h\right)}$ for
$h>0$, which has a mean of $1/\Lambda$) over the total time
baseline, and randomly assign each event an energy drawn from the
assumed distribution. If at least one event falls into any
observing window and has an energy above the LPA sensitivity, the
simulation is counted as a detection. After the $N$ simulations
for a given model, parameter set, and $\Lambda$, we estimate the
nondetection probability $P_{\rm non}\left(\Lambda;\theta\right)$
and find the largest $\Lambda$ such that $P_{\rm
non}\left(\Lambda;\theta\right)>1-99.7\%$. This $\Lambda_{\rm
max}$ is adopted as the $3\sigma$ upper limit on the burst rate
for that $\theta$. The results are displayed as heatmaps for each
model.

\begin{figure}[htbp]
\gridline{\fig{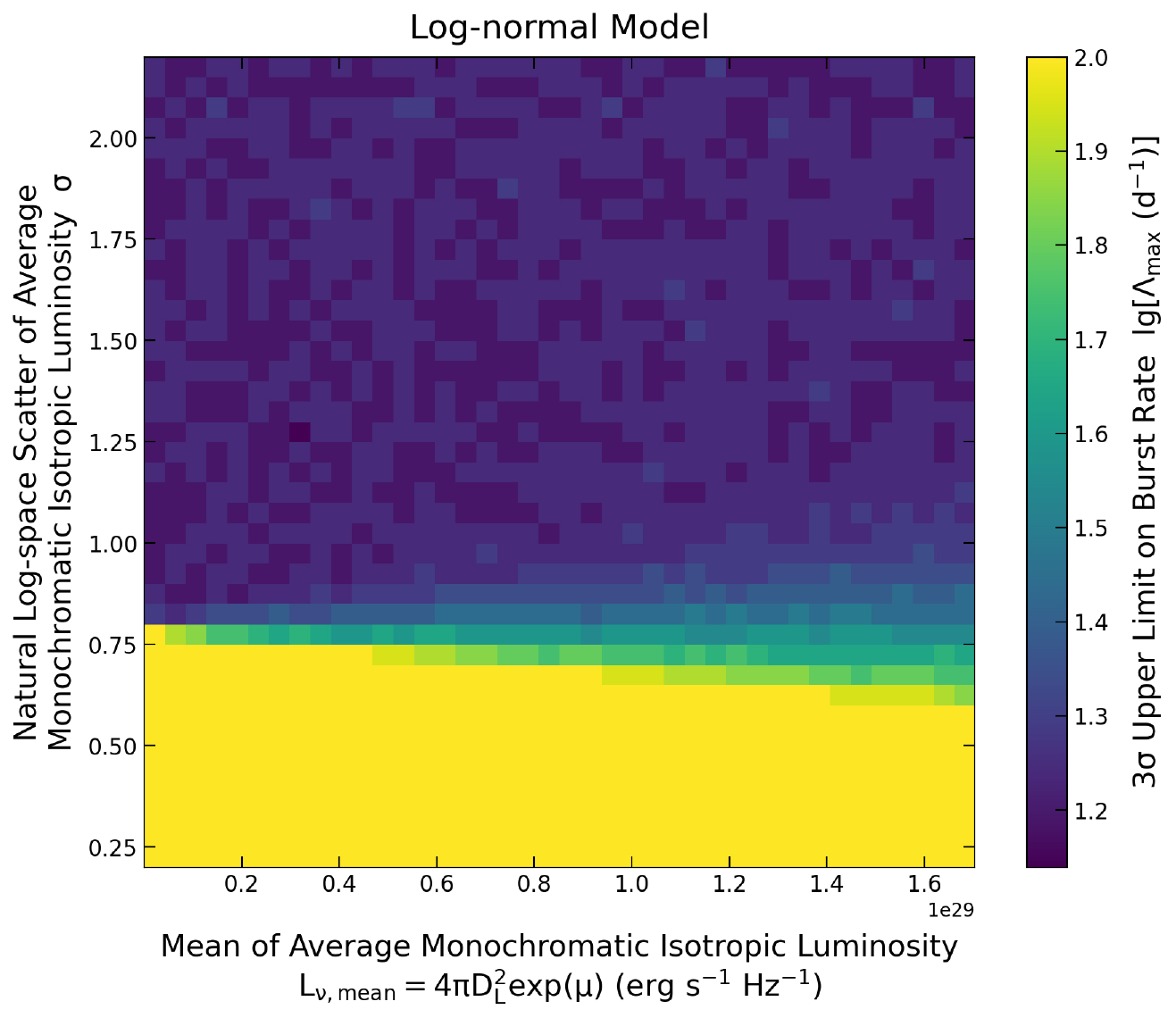}{1\textwidth}{}}
\caption{
Upper limits on the burst rate of FRB-like pulses from SGR 1935+2154 at
$\sim$110 MHz under the log-normal energy distribution model. The heatmap
shows the maximum burst rate $\Lambda_{\rm max}$ (in bursts per day, color
scale) as a function of the mean of the average monochromatic isotropic
luminosity $L_{\nu{\rm ,mean}}$ and the natural log-space scatter $\sigma$ of the
log-normal luminosity distribution. Yellow regions correspond to parameter
combinations for which $\Lambda_{\rm max}$ reaches the upper bound of the
scanned range (${10}^2~{\rm d^{-1}}$). For these parameter combinations,
most bursts drawn from the distribution lie below the sensitivity
threshold, leading to weak constraints.
}
\label{figure8}
\end{figure}

\begin{figure}[htbp]
\gridline{\fig{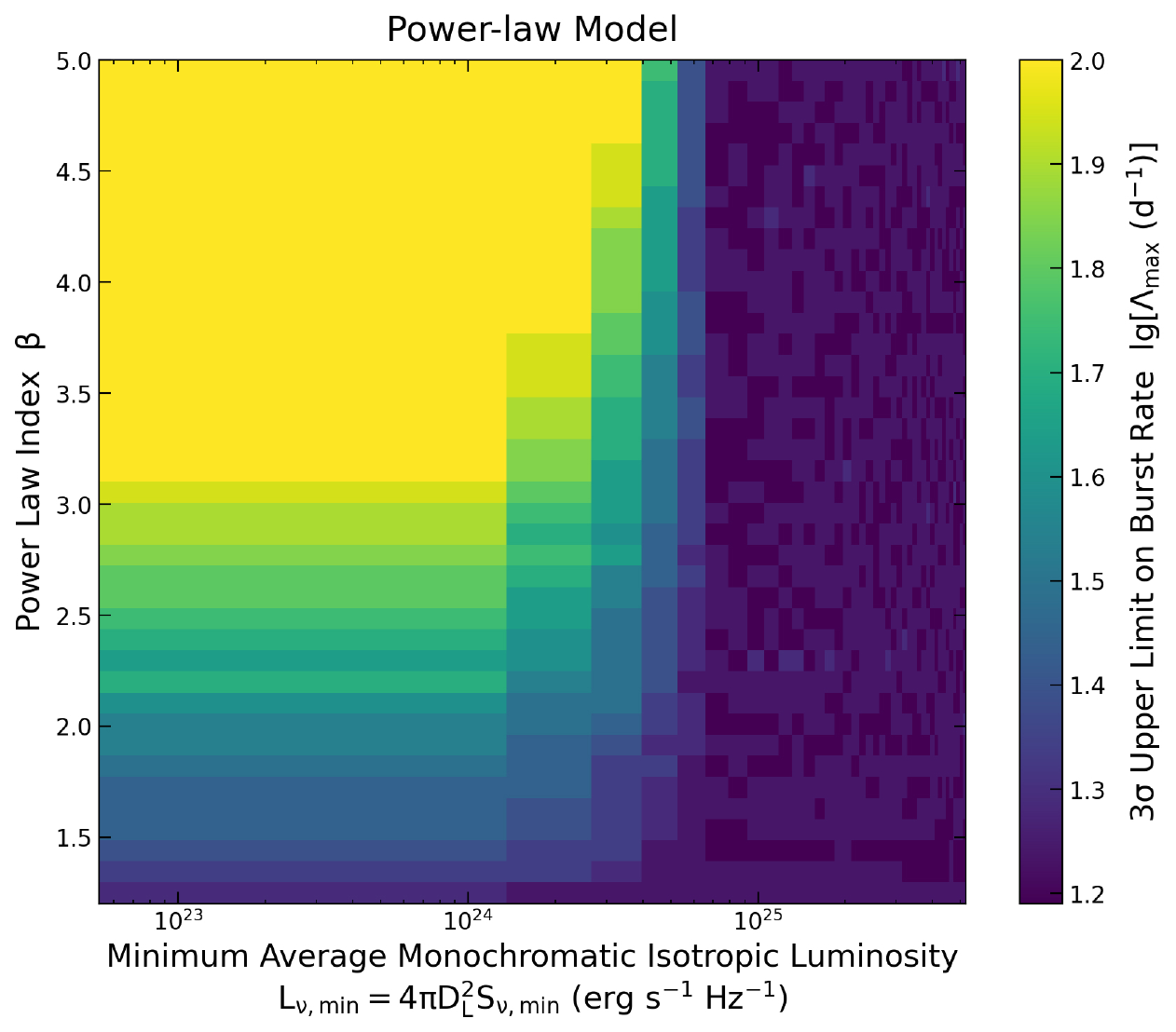}{1\textwidth}{}}
\caption{
Upper limits on the burst rate of FRB-like pulses from SGR 1935+2154 at
$\sim$110 MHz under the power-law energy distribution model. The heatmap
shows $\Lambda_{\rm max}$ (bursts per day, color scale) as a function of
the minimum average monochromatic isotropic luminosity $L_{\nu{\rm ,min}}$
and the power-law index $\beta$. For large $\beta$ and very small
$L_{\nu{\rm ,min}}$ the limits are weak (top of the color scale),
whereas the constraints are stronger elsewhere.
}
\label{figure9}
\end{figure}

Several points deserve emphasis. First, we adopt a frequentist rather than
a Bayesian viewpoint for obtaining the burst-rate upper limits.
Intuitively, a Bayesian would choose the largest $\Lambda$ satisfying
$P_{\rm non}\left(\Lambda;\theta\right)>99.7\%$, which would compress the
allowed $\Lambda$ to extremely small values because only very low burst
rates can guarantee such a high nondetection probability. That would
produce artificially low (non-conservative) limits. In the frequentist
approach, given the observational fact of nondetection, if one claimed a
large $\Lambda$, that claim would be highly improbable; we therefore
reject, at the 99.7\% confidence level, all $\Lambda$ that lead to a
nondetection probability less than 0.3\%. The remaining $\Lambda$ are
acceptable, and we take the largest among them as the upper limit. This
yields a larger (more conservative) allowable range. Second, in each
simulation we assume that any FRB-like pulses from SGR 1935 at $\sim$110
MHz between March and November 2020, if present, can be described by a
stationary Poisson process with a fixed burst rate and a fixed energy
distribution. Third, the telescope sensitivity derived from Equation (\ref{eq9})
is expressed in terms of average flux density, therefore in our MC
realizations the average flux density is the quantity directly simulated.
Since we are ultimately interested in the intrinsic burst energy and
considering that the true pulse width and bandwidth of any potential burst
are unknown, when presenting the heatmaps we convert the distribution
parameters defined in flux density units (i.e., $\mu$ and
$S_{\nu{\rm ,min}}$) into an average monochromatic isotropic luminosity
$L_\nu$  using
\begin{eqnarray}
\label{eq10}
L_\nu=\frac{4{\rm \pi}}{1+z}D_L^2S_\nu,
\end{eqnarray}
where $z=0$ is the redshift of SGR 1935 and $D_L$ is its
luminosity distance. Various $D_L$ has been reported in the
literature. Early works assuming association with the supernova
remnant G57.2+0.8 gave $D_L$ as 9 kpc \citep{2011A&A...536A..83S}
and 12.5 kpc \citep{2018ApJ...852...54K}. Later CO line
observations of G57.2+0.8 yielded $D_L = 6.6$ kpc
\citep{2020ApJ...905...99Z}. After FRB 200428,
\citet{2020ApJ...898L...5Z} used its DM to constrain the distance
as $9.0\pm2.5$ kpc, while \citet{2021MNRAS.503.5367B} combined
multi-frequency observations to give $D_L = 1.5$ --6.5 kpc.
Following \citet{2023SciA....9F6198Z}, we adopt $D_L =6.6$ kpc in
our study, which is an intermediate value among various reports
mentioned above.

The resulting upper limits of the burst rate
are shown in Figures \ref{figure8} and \ref{figure9}.
In the lower part of Figure \ref{figure8} (log-normal model),
for all values of the mean monochromatic isotropic
luminosity ($L_{\nu{\rm ,mean}}$), an excessively small
scatter $\sigma$ leads the upper limit $\Lambda_{\rm max}$ to be close to the upper
bound of our scanned range, ${10}^2~{\rm d^{-1}}$. With a very small
$\sigma$, the energy distribution is concentrated at low energies and
rarely extends above the LPA sensitivity, making almost all bursts
undetectable. As $L_{\nu{\rm ,mean}}$ or $\sigma$ increases, the distribution yields a
higher probability of producing luminous bursts above the LPA sensitivity,
so $\Lambda_{\rm max}$ must be lowered to match the nondetection. For
example, for $L_{\nu{\rm ,mean}}$ between $1.0\times{10}^{29}$ -- $1.6\times{10}^{29}
~{\rm erg~s^{-1}~{Hz}^{-1}}$, with $\sigma\approx0.85$, we have
$\Lambda_{\rm max}\approx{10}^{1.5}~{\rm d^{-1}}$. When $\sigma$ exceeds about $1.15$,
$\Lambda_{\rm max}$ is roughly ${10}^{1.3}~{\rm d^{-1}}$ irrespective of
$L_{\nu{\rm ,mean}}$. Noting that the burst rate estimates presented above, while
seemingly high, simply reflect the fact that most bursts lie below the
sensitivity threshold and only occasional events exceed it. Such MC-based
estimation of the burst-rate upper limits is physically more informative than a simple energy
upper limit derived from the telescope sensitivity, because it quantifies the
burst rates that could be present below the sensitivity threshold.

Figure \ref{figure9} (power-law model) displays $\Lambda_{\rm max}$ for different
$\left(L_{\nu{\rm ,min}},\beta\right)$ combinations. In the left part of
the plot, for large power-law indices $\beta$, high-energy bursts are
rare, so $\Lambda_{\rm max}$ again is near the upper bound of the scanned
range (${10}^2~{\rm d^{-1}}$), indicating a poor constraint because
practically no bursts are detectable. As $\beta$ decreases and the minimum
average monochromatic isotropic luminosity $L_{\nu{\rm ,min}}$ increases,
the probability of exceeding the sensitivity threshold grows, and
$\Lambda_{\rm max}$ becomes well determined. For example, for
$L_{\nu{\rm ,min}}\lesssim0.7\times{10}^{25}~{\rm erg~s^{-1}~{Hz}^{-1}}$
and $\beta\lesssim3.0$, $\Lambda_{\rm max}$ decreases from
$\sim$${10}^{2.0}$ to $\sim$${10}^{1.5}~{\rm d^{-1}}$ as $\beta$ falls. For
$L_{\nu{\rm ,min}}\gtrsim0.7\times{10}^{25}~{\rm erg~s^{-1}~{Hz}^{-1}}$,
$\Lambda_{\rm max}\approx{10}^{1.3}~{\rm d^{-1}}$ regardless of $\beta$.

\subsection{Pulsar-like Pulses}
\label{sec3.2:pulsar-like pulses}

\begin{figure}[htbp]
\gridline{\fig{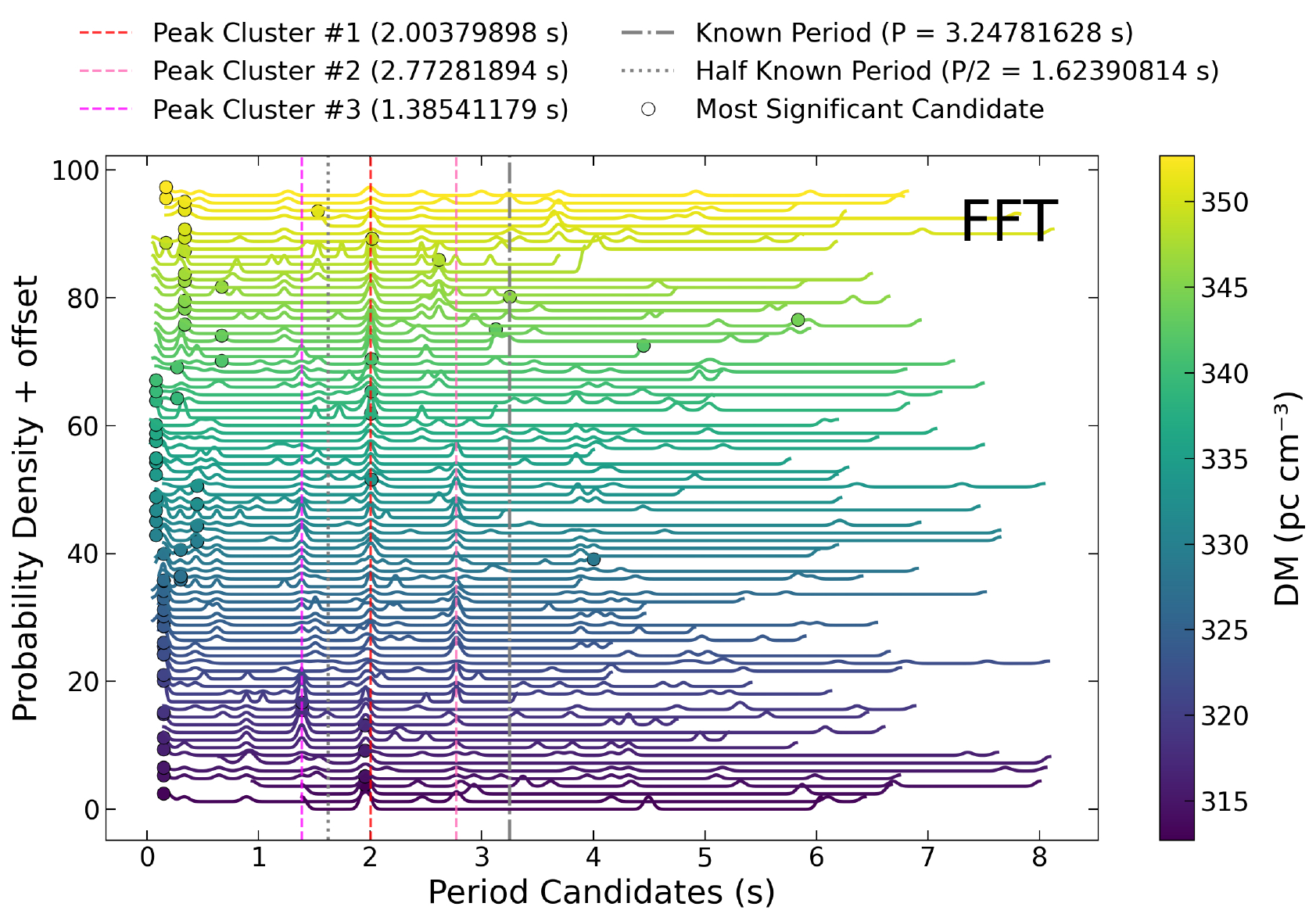}{1\textwidth}{}}
\caption{
FFT periodicity results for pulsar-like emissions. Each colored curve
represents the probability density of candidate periods (from daily
observation segments) after dedispersion with a specific DM using the
Fast Fourier Transform (FFT) method. The most significant candidate for each DM
is marked with a dot. Three vertical ridges (Peak Clusters \#1--3) are
selected as potential periods (colored dash lines). The rotation
period of SGR 1935+2154 (3.24781628 s) and its half value are overplotted (gray
dash-dotted and dotted lines, respectively).
The clustering of dots at the left edge
indicates that the FFT favors short periods because of the finite
observation length.
}
\label{figure10}
\end{figure}

Using the curated LPA 32-channel data of SGR 1935 during
March -- May and September -- November 2020, we performed a pulsar-like
pulse search with our customized strategy. First, we searched for
potential periodicity by using the FFT method,
aiming for periods below 3 s, because FFT searches for longer periods
are plagued by red noise (see Section \ref{sec2.2.1:periodicity
search}). The results are shown in Figure \ref{figure10}, where each
colored curve represents the probability density of
candidate periods obtained after dedispersion with a different DM. For a
genuine periodic signal, it should be detectable over a tolerable DM
mismatch; therefore, vertical ridges formed by overlapping peaks across
different DM curves in Figure \ref{figure10} are potential periods. We
selected the three most densely overlapping ridges as the final periods
for folding (Peak Clusters \#1, 2, 3, marked by colored vertical dash
lines). For harmonic comparison, we also overplot the known period of
SGR 1935 and its half, as measured by FAST during the October 2020
pulsar-like active episode \citep{2023SciA....9F6198Z} .

\begin{figure}[htbp]
\gridline{\fig{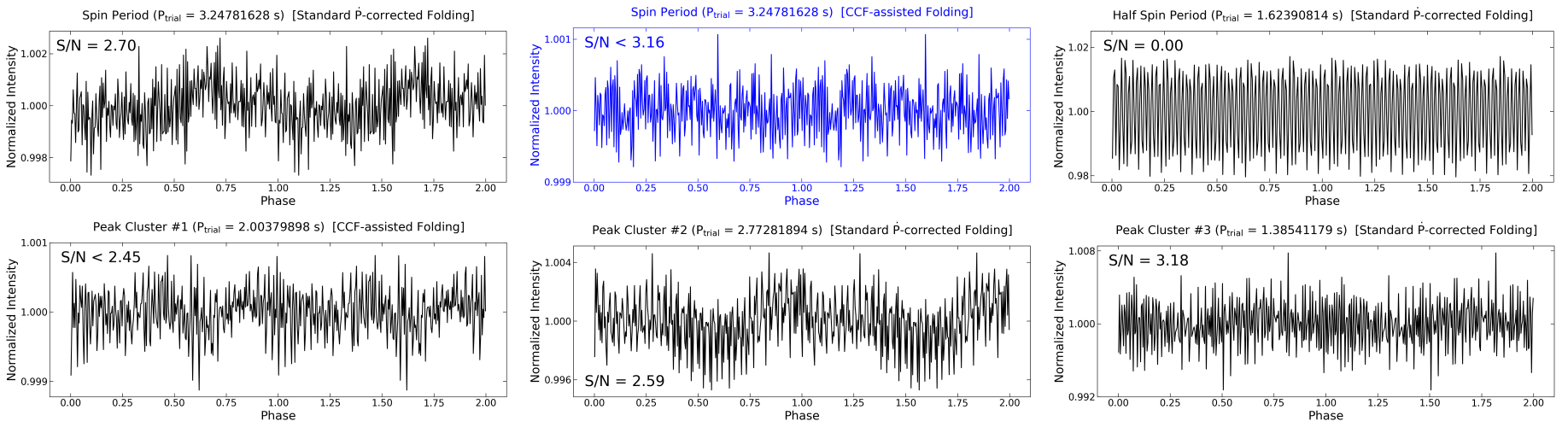}{1\textwidth}{}}
\caption{
Folded profiles obtained with the branched folding strategy, taking
DM = 332.7 $\rm pc~cm^{-3}$. The branched folding strategy, described in
Section \ref{sec2.2.2:CCF-assisted folding method}, adopts CCF-assisted
stacking when more than 50 days exhibit non-flat daily profiles, and a
standard $\dot{P}$-corrected folding otherwise. Black panels correspond
to the profiles assigned by the branched strategy; the blue panel is
additionally provided for completeness and shows the CCF-assisted
stacked profile at the known spin period. First row, from left to right:
standard $\dot{P}$-corrected folding at the spin period of SGR
1935+2154, CCF-assisted stacking at the same period, and folding at half
the spin period. Second row: folded profiles at the three FFT-derived
candidate periods (Peak Clusters \#1--3). All visible peaks in the
folded profiles lie below our periodic-pulse detection threshold of
S/N = 6.
}
\label{figure11}
\end{figure}

\begin{figure}[htbp]
\gridline{\fig{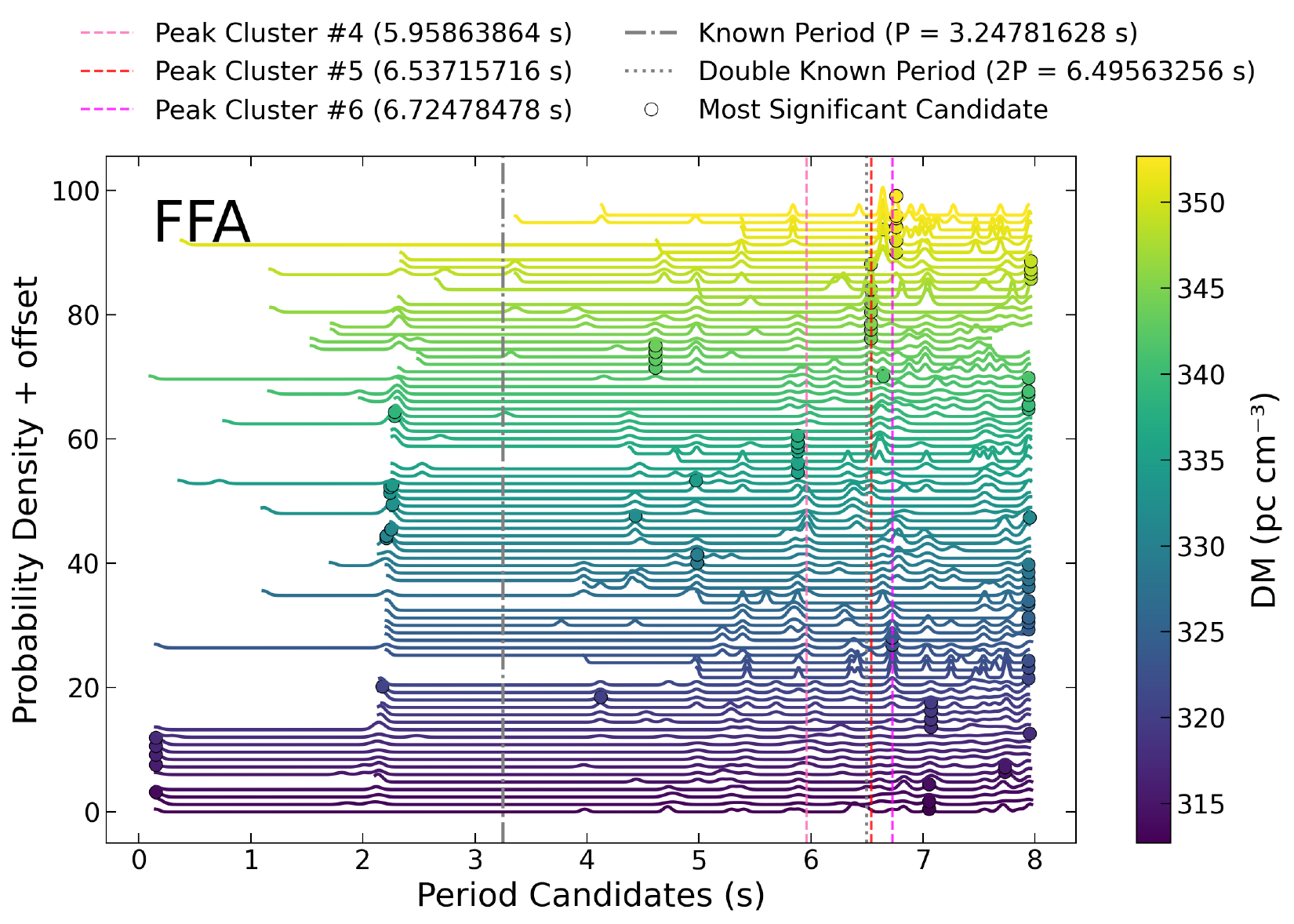}{1\textwidth}{}}
\caption{
Fast Folding
Algorithm (FFA) verification of the candidate periods. The three
short-period candidates (Peak Clusters \#1--3) found by using FFT
are not recovered by the FFA method, confirming that they are
spurious. Instead, the FFA method reveals three longer-period
candidates (Peak Clusters \#4, 5, 6) above 3 s.
}
\label{figure12}
\end{figure}

\begin{figure}[htbp]
\gridline{\fig{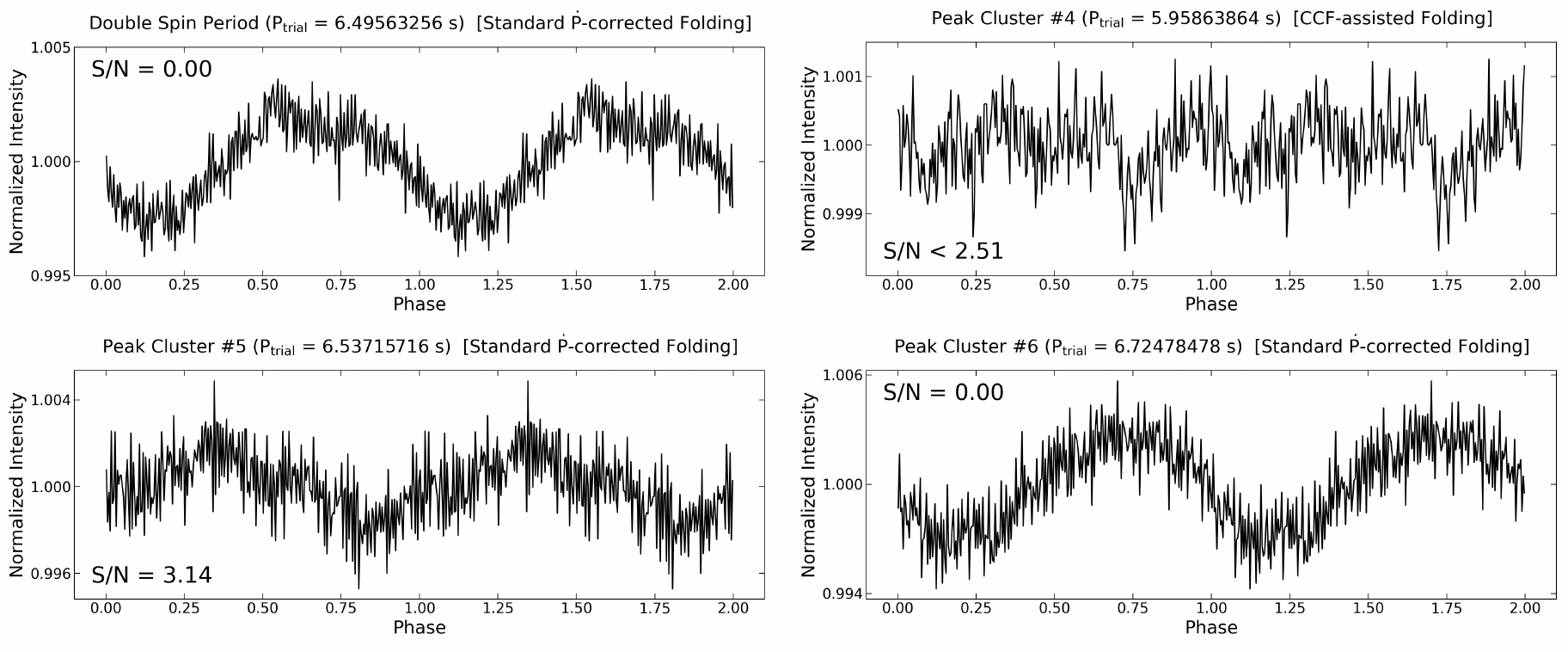}{1\textwidth}{}}
\caption{
Folded profiles of the long-period candidates identified by using FFA
method, obtained with the branched folding strategy described in Section
\ref{sec2.2.2:CCF-assisted folding method}. Folding the curated  LPA
data (dedispersed with DM = 332.7 $\rm
pc~cm^{-3}$) at Peak Clusters \#4--6 and at twice the known
rotation period of SGR 1935+2154 (6.49563256 s) does not produce
any significant pulse peak, indicating the absence of detectable
pulsar-like emission at these periods.
}
\label{figure13}
\end{figure}

For many DM curves, the most significant
candidate (marked with a dot) is notably located at the leftmost end of the
plotted period range, i.e., the lower bound of the FFT search. This
suggests that the candidate periods of Peak Clusters \#1--3 may be spurious.
In the absence of a true period, the finite observation length naturally
samples a larger number of cycles for shorter periods, thereby making them
appear more significant in the FFT power spectrum. Nevertheless, to avoid
missing any possible periodic emission, we folded the data at the true DM
(332.7 $\rm pc~cm^{-3}$) using the candidate periods of Peak
Clusters \#1--3, as well as the known period and its half.
For each of these trial periods, we followed the kurtosis-based
branched folding strategy detailed in Section \ref{sec2.2.2:CCF-assisted
folding method}, where CCF-assisted folding was applied only when more
than 50 days exhibited non-flat daily profiles; otherwise, a standard
$\dot{P}$-corrected folding was adopted on the global time series.
The results are presented in Figure \ref{figure11}.

For the Peak Clusters \#1--3 (second row of Figure \ref{figure11}),
the folded profiles exhibit peaks, but their S/N values are all below
our detection threshold of S/N = 6 for periodic pulses; therefore, these
features cannot be distinguished from random noise fluctuations. We note
that the trial period corresponding to Peak Cluster \#3 is 1.38541179 s,
which is close to, but clearly distinct from, the known period of
J1921+2153 ($\sim$1.34 s). Any residual weak signal from J1921+2153
would be completely smeared out by the large DM mismatch (12.4 vs. 332.7
$\rm pc~cm^{-3}$) when the data are dedispersed at the DM of SGR 1935.
The detectable contaminating pulses of J1921+2153 on the four affected
days were already masked through a low-DM single-pulse pre-search, and
any residual weak pulses that escaped the pre-search would be buried in
the noise and completely smeared out by the large DM mismatch when the data are dedispersed at the DM of SGR
1935. Furthermore, this trial period of 1.38541179 s was assigned to the
standard $\dot{P}$-corrected folding, and the use of SGR 1935's
$\dot{P}$ in folding would cause additional phase drift over the
nine-month baseline. We can therefore safely rule out any physical
pathway for J1921+2153 contamination at this trial period; the weak peak
is more plausibly of noise origin.

No pulse peak
is detected in the folded profile at half the known rotation period
(first row, third column of Figure \ref{figure11}).
The folded profile at the rotation period (first row, first
column of Figure \ref{figure11}) does display a peak, but its S/N
is 2.70. Although the branched folding strategy assigned the standard
$\dot{P}$-corrected folding to the know rotation period, we also show
the CCF-assisted stacked profile for completeness (first row, second
column of Figure \ref{figure11}); it also exhibits a sharp peak, with its
S/N below 3.16 (the CCF-assisted folding only provides an upper limit
to the true S/N), which still remains below our periodic-pulse detection
threshold of S/N =6. Hence we cannot
claim unequivocally that pulsar-like radiation from SGR 1935 exists at
$\sim$110 MHz, but the presence of a weaker signal cannot be excluded.

To complement the FFT-based periodicity search, which is limited
to periods $\lesssim$3 s, we also performed an FFA search. The FFA is
less susceptible to red noise and not restricted to Fourier frequency
bins, making it well suited for secondary verification, particularly for
longer periods. As shown in Figure \ref{figure12},
the three candidate periods (Peak Clusters \#1--3) found by the
FFT were not recovered by the FFA, confirming that they were
indeed spurious, which corroborates our initial assessment. For
periods longer than 3 s, the FFA revealed three new candidate
periods, labelled Peak Clusters \#4, 5, 6. Among these, Peak
Cluster \#5 (6.53715716 s) is close to twice of the spin period of
SGR 1935 (6.49563256 s). To check for a possible harmonic
relation, we folded the data at twice the spin period, as well as at the
candidate periods themselves; none of these produced a significant pulse
peak (Figure \ref{figure13}).

We have tried to derive an upper limit on the energy of any possible
pulsar-like emission by using the LPA sensitivity, which is calculated as
\citep{2012hpa..book.....L, 2022MNRAS.517.1112T}
\begin{eqnarray}
\label{eq11}
\mathcal{S}_{\rm pp}=\frac{\left({\rm S/N}\right)_{\rm min,pp}T_{\rm sys}}
{G\sqrt{N_{\rm pol}\Delta\mathcal{V}t_{\rm tot}}}\sqrt{\frac{W_{\rm obs}}
{P-W_{\rm obs}}},
\end{eqnarray}
where $\left({\rm S/N}\right)_{\rm min,pp}=6$ is the adopted
periodic-pulse detection threshold, $t_{\rm tot}=31432~{\rm s}$ is the
total effective observing time on
SGR 1935 during March -- May and September -- November 2020, and
$P=3.24781628~{\rm s}$ is the rotation period measured in October
2020 \citep{2023SciA....9F6198Z}. The observed pulse width $W_{\rm obs}$
(if it exists) remains uncertain; to obtain a conservative upper limit we
adopt a typical duty cycle $W_{\rm obs}/\left(P-W_{\rm obs}\right)=0.1$
\citep{2012hpa..book.....L}. Using the same parameters as in Equation (\ref{eq9}),
we find $\mathcal{S}_{\rm pp}=0.40~{\rm mJy}$. Again, lacking knowledge of
the true pulse width and bandwidth, we convert this flux density to an
average monochromatic isotropic luminosity via Equation (\ref{eq10}),
yielding $2.08\times{10}^{19}~{\rm erg~s^{-1}~{\rm Hz}^{-1}}$.


\newpage

\section{Summary}
\label{sec4:summary}

In this study, we reanalyzed the LPA observations of the Galactic
FRB source SGR 1935+2154 during its two active periods in 2020
(March -- May and September -- November) using customized search
methods for both FRB-like and pulsar-like signals. Our main
results are summarized below:\

(1) No significant FRB-like pulses were detected. $3\sigma$ upper
limits on the burst rate at $\sim$110 MHz were derived through
MC simulations. Under a log-normal energy distribution, the upper
limit is $\sim$${10}^{1.5}~{\rm d^{-1}}$ for a mean of average
monochromatic isotropic luminosity $L_{\nu{\rm
,mean}}\sim1.3\times{10}^{29} ~{\rm erg~s^{-1}~{Hz}^{-1}}$ and a
natural log-space scatter $\sigma\sim0.85$. For a power-law distribution, the
limit is $\sim$${10}^{1.8}~{\rm d^{-1}}$ for an index
$\beta\lesssim3.0$ and a minimum average monochromatic isotropic
luminosity $L_{\nu{\rm ,min}}\lesssim0.7\times{10}^{25} ~{\rm
erg~s^{-1}~{Hz}^{-1}}$. These constraints are more informative
than a simple energy upper limit because they quantify the burst
rates that could exist below the sensitivity threshold.\

(2) No pulsar-like pulses were detected either. The data folded
at the known spin period of SGR 1935+2154 (3.24781628 s) using both the
standard $\dot{P}$-corrected folding and the CCF-assisted stacking
yielded a weak pulse peak, with S/N = 2.70 and < 3.16, respectively;
both values lie below our periodic-pulse detection threshold of S/N = 6,
and thus neither can be regarded as a secure detection.
However, we cannot rule out the presence of a
weak periodic signal from currently available observational data.
A conservative upper limit on the average monochromatic isotropic
luminosity of such pulsar-like emission was derived as
$2.08\times{10}^{19} ~{\rm erg~s^{-1}~{Hz}^{-1}}$.

In summary, we find no clear evidence for either FRB-like or
pulsar-like emission from SGR 1935+2154 at $\sim$110 MHz. Our
results place stringent low-frequency constraints on the burst
rate and luminosity of this prototypical Galactic FRB source. The
weak hint of a periodic signal at the known rotation period,
albeit not statistically significant, leaves open the possibility
of very faint pulsar-like radiation at meter wavelengths.



\clearpage

\section*{Acknowledgements}
We are grateful to the anonymous referee for useful comments and 
suggestions. Shi-Jie Gao and Guanhong Lin are acknowledged for valuable discussions.
This study is supported by the National Natural Science Foundation
of China (Grant No. 12233002), by the National Key R\&D Program of China
(2021YFA0718500). YFH also acknowledges the support from the Xinjiang
Tianchi Program. EAB and SAT were supported by the Russian Science
Foundation No. 22-12-00236-\CYRP. The work of GET was carried out within
the framework of the state assignment of the Keldysh Institute of Applied
Mathematics of the Russian Academy of Sciences (No. FFMN-2025-0024).
JJG acknowledges support from the Youth Innovation Promotion Association (2023331).
OA was also supported by the Project funded by China Postdoctoral Science
Foundation (Grant No. 2025M783225).


\bibliography{references}{}
\bibliographystyle{aasjournal}

\end{CJK*}
\end{document}